\documentclass[12pt]{article}
\pdfoutput=1
\usepackage{epsfig,cite}
\usepackage {amsmath}
\usepackage{amssymb}
\include{epsf}
\newcount\timecount
\newcount\hours \newcount\minutes  \newcount\temp \newcount\pmhours
\hours = \time
\divide\hours by 60
\temp = \hours
\multiply\temp by 60
\minutes = \time
\advance\minutes by -\temp
\def\hour{\the\hours}
\def\minute{\ifnum\minutes<10 0\the\minutes
            \else\the\minutes\fi}
\def\clock{
\ifnum\hours=0 12:\minute\ AM
\else\ifnum\hours<12 \hour:\minute\ AM
      \else\ifnum\hours=12 12:\minute\ PM
            \else\ifnum\hours>12
                 \pmhours=\hours
                 \advance\pmhours by -12
                 \the\pmhours:\minute\ PM
                 \fi
            \fi
      \fi
\fi
}

\def\monthname{\relax\ifcase\month 0/\or January\or February\or
   March\or April\or May\or June\or July\or August\or September\or
   October\or November\or December\else\number\month/\fi}

\def\bold#1{\setbox0=\hbox{$#1$}%
     \kern-.025em\copy0\kern-\wd0
     \kern.05em\copy0\kern-\wd0
     \kern-.025em\raise.0433em\box0 }

\def\beq{\begin{equation}}
\def\eeq{\end{equation}}
\def\beqn{\begin{equation}}
\def\eeqn{\end{equation}}
\def\neq{n_{\rm eq}}

\def\tcm{\theta_{\rm\scriptscriptstyle CM}}

\def\ga{\mathrel{\raise.3ex\hbox{$>$\kern-.75em\lower1ex\hbox{$\sim$}}}}
\def\la{\mathrel{\raise.3ex\hbox{$<$\kern-.75em\lower1ex\hbox{$\sim$}}}}
\def\gev{{\rm \, Ge\kern-0.125em V}}
\def\tev{{\rm \, Te\kern-0.125em V}}
\def\gyr{{\rm \, G\kern-0.125em yr}}
\def\ohsq{\Omega_{\chi} h^2}

\def\tsq{|{\cal T}|^2}

\def\tb{\tan \beta}

\def\gappeq{\mathrel{\rlap {\raise.5ex\hbox{$>$}}
{\lower.5ex\hbox{$\sim$}}}}
\def\lappeq{\mathrel{\rlap{\raise.5ex\hbox{$<$}}
{\lower.5ex\hbox{$\sim$}}}}
\def\Toprel#1\over#2{\mathrel{\mathop{#2}\limits^{#1}}}

\def\avg#1{\left\langle #1 \right\rangle}

\def\m12{m_{1\!/2}}

\def\bea{\begin{eqnarray}}
\def\eea{\end{eqnarray}}

\begin{document}
\begin{titlepage}
\pagestyle{empty}
\baselineskip=21pt
\rightline{KCL-PH-TH/2015-12, LCTS/2015-04, CERN-PH-TH/2015-049}
\rightline{UMN--TH--3424/15, FTPI--MINN--15/10}
\vskip 0.75in
\begin{center}
{\large{\bf Gluino Coannihilation Revisited}}
\end{center}
\begin{center}
\vskip 0.3in
{\bf John~Ellis}$^{1,2}$,
{\bf Feng~Luo}$^2$
and {\bf Keith~A.~Olive}$^{3,4}$
\vskip 0.3in
{\small {\it
$^1${Theoretical Particle Physics and Cosmology Group, Department of
  Physics, \\ King's College London, London WC2R 2LS, United Kingdom}\\
$^2${Theory Division, CERN, CH-1211 Geneva 23, Switzerland}\\
$^3${School of Physics and Astronomy, University of Minnesota, Minneapolis, MN 55455, USA}\\
$^4${William I. Fine Theoretical Physics Institute, School of Physics and Astronomy,\\
University of Minnesota, Minneapolis, MN 55455, USA}\\
}}

\vskip 0.75in
{\bf Abstract}
\end{center}
\baselineskip=18pt \noindent

Some variants of the MSSM feature a strip in parameter space where the lightest neutralino $\chi$
is identified as the lightest supersymmetric particle (LSP), the gluino
${\tilde g}$ is the next-to-lightest supersymmetric particle (NLSP) and is nearly degenerate with $\chi$, and the relic
cold dark matter density is brought into the range allowed by astrophysics and cosmology 
by coannihilation with the gluino NLSP. We calculate the relic density along this
gluino coannihilation strip in the MSSM, including the effects of gluino-gluino bound states
and initial-state Sommerfeld enhancement, and taking into account the decoupling of the gluino
and LSP densities that occurs for large values of the squark mass
$m_{\tilde q}$. We find that bound-state effects can increase the
maximum $m_\chi$ for which  the relic cold dark matter density lies within the range
favoured by astrophysics and cosmology by as much as $\sim 50$\% if $m_{\tilde q}/m_{\tilde g} = 1.1$,
and that the LSP may weigh up to $\sim 8$~TeV for a wide range of $m_{\tilde q}/m_{\tilde g} \lesssim 100$.

\vfill
\leftline{March 2015}
\end{titlepage}
\baselineskip=18pt

\section{Introduction}

In the absence of any signal for supersymmetry during Run~1 of the LHC \cite{lhc}, it is
natural to ask how and where supersymmetry may be hiding. Perhaps it is hiding
in plain sight with a compressed spectrum \cite{compress} that the conventional missing-energy
searches at the LHC have been unable to resolve? Or perhaps $R$ parity is violated,
in which case supersymmetry may be hiding among the jets and leptons produced
by Standard Model processes? Or perhaps $R$ parity is conserved, but supersymmetric particles are too heavy
to have been detected during Run~1 of the LHC?

There are two issues with this last possibility. One is the accentuation of the
problem of the naturalness (or fine-tuning) of the electroweak scale that low-scale
supersymmetry was postulated to mitigate, and the other is the cosmological cold
dark matter density. The cold dark matter may well not consist only, or even predominantly,
of the lightest supersymmetric particle (LSP). However, even if the cold dark matter density is
considered only as an upper limit on the relic LSP density, it imposes an upper bound on
the LSP mass that depends on the specific LSP candidate under consideration.

If $R$ parity is conserved and the LSP is present in the Universe today as a relic
from the Big Bang, it is expected to be electromagnetically neutral and have only
weak interactions. In the minimal supersymmetric extension of the Standard Model (MSSM),
the most plausible candidates are the lightest neutralino $\chi$ and the gravitino  \cite{ehnos}. Here we
consider the neutralino case, and the cosmological upper bound on its mass.

The relic LSP density depends not only on the LSP mass, but also on the
rates at which it annihilated with itself and coannihilated with other sparticles
in the early Universe \cite{gs}. Other things being equal, the largest LSP mass is allowed
when such coannihilation rates are maximised, which happens when the LSP is
(nearly) degenerate with other particles. If there is only one such coannihilating sparticle
species, the coannihilation rate will in general be maximised for a coloured sparticle.
There have been analyses in the literature of the cases where the coannihilating
particle is a squark, specifically the lighter stop squark \cite{stop,eos,deos,Harz,EOZ,raza,ibarra}, and also the case of the gluino
\cite{pro,nath,shafi,hari,gluino2014,deSimone:2014pda,liantao,raza}.
In general, one would expect that the heaviest LSP will be allowed when it coannihilates with the particle
with the largest colour charge, namely the gluino.

We study here the question how heavy the neutralino LSP $\chi$ could
be, if it is nearly degenerate with, and coannihilates with, the gluino ${\tilde g}$. This is
of relevance to assessing, for example, what centre-of-mass energy would be needed
for a proton-proton collider to be `guaranteed' to detect $R$-conserving supersymmetry.
There can of course be no cast-iron guarantee, even within the MSSM. For example,
even in the gluino coannihilation case studied here the neutralino LSP mass limit depends on the
squark masses, and the LSP mass limit could be substantially modified if the squarks were degenerate
with the neutralino LSP and the gluino. However, a complete analysis of this case lies beyond the scope
of this paper.

As already mentioned, there have been several previous analyses of neutralino-gluino
coannihilation \cite{pro,nath,shafi,hari,gluino2014,deSimone:2014pda,liantao,raza}, and the main new elements here are in our discussions of the effects of
gluino-gluino bound states and of the issue whether coannihilations can be maintained in the presence of a large squark to gluino mass ratio. Here, we will restrict our attention to the coannihilation processes
and leave their application to more complete models (with for example, radiative electroweak 
symmetry breaking) for future work \cite{eelo}.
As we discuss in detail, bound states can remove from the primordial plasma gluino pairs
that may subsequently annihilate into Standard Model particles, before they can decay
into the LSP as is usually assumed in discussions of coannihilation.
We present numerical estimates of the bound-state production rate,
and find that, for fixed sparticle masses, the relic dark matter density is substantially reduced compared with the cases
where bound-state formation is neglected.
Conversely, the cosmological relic density may lie within the cosmological range for substantially {\it larger} LSP masses
than would have been estimated in the absence of bound-state effects: this effect is $\sim 50$\% for
$m_{\tilde q}/m_{\tilde g} = 1.1$, falling to $\sim 20$\% for $m_{\tilde q}/m_{\tilde g} \sim 10$ to $50$.
Another effect we discuss is that, if $m_{\tilde q}/m_{\tilde g} \ga 100$, the densities of neutralinos and
gluinos decouple and coannihilation effects
freeze-out early, leaving a significantly higher relic density, thereby {\it reducing} the possible LSP mass.
There is also a reduction in the possible LSP mass for small $m_{\tilde q}/m_{\tilde g} \to 1$, due to
cancellations between $s$-, $t$- and $u$-channel diagrams that tend to reduce annihilation rates.

Taking these effects into account, we find a maximum LSP mass $\sim 8$~TeV if it is the Bino, which
may be attained for $10 \la m_{\tilde q}/m_{\tilde g} \la 100$. If the LSP is the neutral Wino, the
upper limit is reduced to $\sim 7$~TeV, and for a neutral Higgsino the upper limit it becomes $\sim 6$~TeV.

The layout of this paper is as follows. In section~2, we review the Sommerfeld enhancement in
the relevant gluino-gluino annihilation processes. In Section~3, we discuss the formation of
gluino-gluino bound states, considering also dissociation processes in the early Universe.
In Section~4, we consider the rates for conversion between gluinos and neutralinos.
In Section 5, we present and discuss the coupled Boltzmann equations for neutralinos $\chi$,
gluinos and gluino-gluino bound states.
Section~6 contains some numerical results
for the gluino coannihilation strip and a discussion of its endpoint. Section~7
summarises our conclusions and discusses their significance for future colliders.
Finally, Appendices present technical aspects of the computation of the
$2 \to 2$ cross sections needed for solving the Boltzmann equations.

\section{Sommerfeld Enhancement}

Before discussing the formation and effects of gluino-gluino bound states, we first discuss
briefly Sommerfeld effects in gluino-gluino annihilation, which may enhance
annihilation rates at low velocities, and are particularly relevant in the case of the strongly-interacting gluino. 
As a general rule, initial-state interactions modify s-wave cross-sections by factors \cite{as,somm}
\begin{equation}
F(s) \; \equiv \frac{- \pi s}{1 - e^{\pi s}} : \; s \; \equiv \; \frac{\alpha}{\beta} \, ,
\label{Sommerfeld}
\end{equation}
where $\alpha$ is the coefficient of a Coulomb-like potential whose sign convention is such that
the attractive case has $\alpha < 0$, and $\beta$ is the velocity of one of the annihilating particles in the
centre-of-mass frame of the collision. In the cases of strongly-interacting particles, 
the Coulomb-like potential has the form \cite{coul}
\begin{equation}
V \; = \; \frac{\alpha_s}{2 r} \left[ C_f - C_i - C_i^\prime \right] \, ,
\label{Casimirs}
\end{equation}
where $\alpha_s$ is the strong coupling strength,
$C_f$ is the quadratic Casimir coefficient of a specific final-state colour representation,
and $C_i$ and $C_i^\prime$ are the quadratic Casimir coefficients of the annihilating coloured
particles. In the case of octet annihilating particles such as gluinos, $C_i = C_i^\prime = C_8 = 3$.
The relevant final states are in singlet, octet, or $27_s$ representations, for which
$C_f = 0$, $C_f =3$, or $C_f = 8$. 

As discussed in~\cite{EOZ}, Sommerfeld effects such as these have been implemented in the
{\tt SSARD} code~\cite{SSARD} for calculating the relic
dark matter density. In the coannihilation region of interest, this code uses
a non-relativistic expansion for annihilation cross-sections:
\begin{equation}
\langle \sigma v \rangle \; = \; a + b x^{-1} + \dots \, ,
\label{ab}
\end{equation}
where $\langle ... \rangle$ denotes an average over the thermal distributions of the annihilating
particles, the coefficient $a$ and $b$ represent the contributions of the s- and p-wave cross-sections,
$x \equiv m/T$, and the dots represent terms of higher order in $1/x$. A Sommerfeld enhancement
occurs when $\alpha < 0$ in (\ref{Sommerfeld}), modifying the leading term in (\ref{ab}) so that it
acquires a singularity $\propto \sqrt{x}$. In this paper
we have included these enhancements in the $\tilde g \tilde g \to g g$ and $\tilde g \tilde g \to q {\bar q}$
cross sections. 
The procedure for obtaining a thermally averaged cross section
is given in Appendix A.  The expressions for the matrix elements for the coannihilation processes 
are given in detail in Appendix B.

\section{Gluino-Gluino Bound-State Formation}

Gluino-neutralino coannihilations may increase the effective annihilation cross section
and thereby lower the final neutralino relic abundance.  The Sommerfeld enhancement discussed above
further increases the cross section in specific channels and again lowers the abundance of neutralinos allowing for larger
masses at the tip of the coannihilation strip defined by $\Delta m = 0$ where $\Delta m$ is the gluino-neutralino mass difference
\cite{deSimone:2014pda}.
Gluino-gluino bound states can further serve to remove gluinos from the thermal bath and thereby lower the relic density
by a factor that is non-negligible relative to the Sommerfeld enhancement, and much larger than the
uncertainty in the cosmological cold dark matter density. 

The dominant process for the formation and dissociation of gluino-gluino bound states $\tilde R$ in the thermal plasma 
is ${\tilde g} +{\tilde g} \leftrightarrow {\tilde R} + g$. These processes become important when the plasma temperature falls low enough
for typical thermal energies to become comparable to the binding energy of the $\tilde R$ state, namely
$T \lesssim E_B \equiv 2 m_{\tilde g} - m_{\tilde R}$. In principle, one may form colour-octet states as well as singlets, but the
latter are expected to be more deeply bound with larger wave functions at the origin. Here we focus on the production of
the lightest colour-singlet state, $1_s$, with orbital angular momentum $L=0$ and spin angular momentum $S=0$, which
is expected to be the most copiously produced. Since we are considering gluinos weighing several TeV,
we expect the leading order of QCD perturbation theory to be a useful approximation, and assume the Coulomb potential
$V (r) = -3 \alpha_s / r$ for the $1_s$ state, with binding energy $E_B \simeq (3 \alpha_s/2)^2 m_{\tilde g}$.
The normalised spatial part of the wave function for this $1_s$ bound state is
\beq
\phi_{bs} (r) = (\pi a^3)^{-1/2} e^{-r/a} \, ,
\label{eq:bswavefunction}
\eeq
where $a \equiv 2/(3 \alpha_s m_{\tilde g})$ is the Bohr radius. The $1_s$ bound state decays predominantly to a pair of gluons and the leading order decay rate is 
\beq
\Gamma_{\tilde R} = \frac{243}{4} \alpha_s^5 m_{\tilde g} .
\eeq

\subsection{Dissociation}

In order to calculate bound-state formation and dissociation via the dominant processes
$\tilde{g}^a + \tilde{g}^b \leftrightarrow \tilde{R} + g^c$,
we first calculate the bound-state dissociation cross section, $\sigma_{dis}$
following Section 56 of~\cite{BLP}, where the photoelectric effect for an atom is calculated.
The central part of the calculation is the evaluation of the transition amplitude given in Eq.~(56.2) of~\cite{BLP}:
\beq
\mathcal{M}_{fi} = \int \phi_f^\ast (-i \frac{\vec{\nabla} \cdot \vec{\epsilon}^c}{(m_{\tilde g}/2)}) e^{i \vec{k} \cdot \vec{r}} \phi_i d^3 \vec{r} \, ,
\label{eq:mfi}
\eeq
where $\phi_f$ is the wave function of the free $\tilde{g}^a$ $\tilde{g}^b$ pair and
$\phi_i \equiv \phi_{bs} (r)$, and $\vec{\epsilon}^c$ and $\vec{k}$ are the polarisation and momentum
vectors of the gluon, respectively. 

We use the dipole approximation, $e^{i \vec{k} \cdot \vec{r}} \approx 1$, which is justified because the bound-state
wave function $\phi_{bs} (r)$ is exponentially suppressed for $r > a$, and because
the gluon momentum $ |\vec{k}| = \omega$, where its energy $\omega$ satisfies the conservation condition
\beq
\omega + {\omega^2 \over 2 (2m_{\tilde g} - E_B)} = {|\vec{p}|^2 \over m_{\tilde g}} + E_B \, ,
\label{conservation}
\eeq
where $|\vec{p}|$ is the momentum of one of the annihilating gluinos. (Note that $|\vec{p}|$ is the same as the
relative momentum, $(m_{\tilde g}/2) v_{rel}$, and  the second term on the LHS of (\ref{conservation}) can be
neglected.) We find $\omega a \simeq E_B a = {(3 \alpha_s) \over 2} \ll 1$ for $v_{rel} = 0$ and $\alpha_s = 0.1$,
and more generally $\omega a < 1$ when $v_{rel} < 0.6$, so that the dipole approximation should be
sufficient for our purposes.

The dipole approximation imposes a selection rule on $\phi_f$, which needs to be in an $L=1$ state. 
Further, charge conjugation ($C$-parity) conservation requires that $C(\tilde{g}^a \tilde{g}^b) =  C(\tilde{R}) C(g^c)$, where
the $1_s$ ground state  with $L=0$ and $S=0$ has $J^{PC} = 0^{-+}$. 
The $C$-parity of the colour anti-symmetric $8_A$ state is the same as that of the gluon ~\cite{Kauth2011},
while the $C$-parity of colour-symmetric $8_s$ state is opposite of that of the gluon, for all color indices.
Therefore, the only possible state for $\phi_f$ is $8_A$, with $L=1$ and $S=0$.
(Note also that parity is conserved in this case, because $P(\phi_f) = 1$ and the gluon has $P = -1$.)

The normalised spatial part of the wave function for the free pair $\tilde{g}^a \tilde{g}^b$ is
\beq
\phi_f = {1 \over 2 |\vec{p}|} {\sum_{L=0}^{\infty}} i^L (2 L + 1) e^{-i \delta_L} R_{pL} (r) P_L ({\vec{p} \cdot \vec{r} \over |\vec{p}| r}) \, . 
\eeq
Only the $L=1$ term survives, due to the selection rule from the dipole approximation.
Since we wish to calculate $|\mathcal{M}_{fi}|^2$, we may discard the phase shift factor
$e^{-i \delta_L}$ ($\delta_L$ is real) and the factor $i^{L}$. Therefore we write  
\beq
\phi_f = {3 \over 2 |\vec{p}|} R_{p1} (r) P_1 ({\vec{p} \cdot \vec{r} \over |\vec{p}| r}) \, . 
\label{eq:freestatewavefunction}
\eeq 
so that 
\beq
d\sigma^0_{dis} = \alpha_s {(m_{\tilde g}/2) |\vec{p}| \over 2 \pi \omega} |\mathcal{M}_{fi}|^2 d \Omega_{\vec{p}} \, ,
\eeq
where $\mathcal{M}_{fi}$ is calculated following Section 56 of~\cite{BLP}.

Since $\phi_f$ is the wave function for an $8_A$ state, the Coulomb potential is $V_f (r) = -{3 \over 2} \alpha_s /r$,
whereas $\phi_i$ is a wave function for the Coulomb potential $V (r) = -3 \alpha_s /r$, the result is different from
Eq. (56.12) of~\cite{BLP}, namely
\beq
\sigma^0_{dis} = {2^9 \pi^2 \over 3} \alpha_s a^2 \left({E_B \over \omega}\right)^4 {(1+\xi^2) \over [1+(\kappa \xi)^2]} {e^{-4 \xi \arctan(1/\kappa\xi)} \over 1-e^{-2 \pi \xi}} \kappa^{-1} \, ,
\label{eq:sigma_dis_wo_factors}
\eeq
where $\xi = {3 \over 2} \alpha_s / v_{rel}$ and $\kappa = 2$. This equation is averaged over the gluon polarizations
and would reduce to Eq. (56.12) if $\kappa = 1$.

The total wave functions for the free $\tilde{g}^a$ $\tilde{g}^b$ pair and the bound state $\tilde{R}$
are products of the spin, colour and spatial parts of the wave functions.
In view of the Majorana nature of the gluinos, the total wave functions should be anti-symmetric. 
Concerning the spin part of the wave function, since both the bound state and the free gluino pair are in
an $S=0$ state, the spin wave functions are both 
\beq
(\uparrow \downarrow - \downarrow \uparrow)/\sqrt{2}  \, ,
\eeq  
and the spin parts of the wave functions do not introduce any extra factor in $\sigma_{dis}$. 
As for the colour part of the wave function, according to~\cite{GoldmanHaber,hy} $\phi_{i, color} = \delta_{de} / \sqrt{8}$,
and $\phi_{f, color} = f_{hae} /\sqrt{3}$ (the latter because $f_{abc} f_{abd} = 3 \delta_{cd}$). 

The $(-i {\vec{\nabla} \cdot \vec{\epsilon}^c \over(m_{\tilde g}/2)} ) e^{i \vec{k} \cdot \vec{r}}$ factor
in the transition amplitude (\ref{eq:mfi}) is calculated from the gluino-gluino-gluon interaction Lagrangian 
\beq
\mathcal{L} = {i \over 2} g_s g^c_\mu f_{abc} \bar{\tilde{g}}^a \gamma^\mu \tilde{g}^b \, ,
\label{QCD}
\eeq
which can be compared with the corresponding QED interaction Lagrangian
\beq
\mathcal{L} = -eQ_f A_\mu \bar{f} \gamma^\mu f \, .
\eeq
We simply replace the electric charge factor $Q_f$ in the transition amplitude (\ref{eq:mfi})
by the colour factor $f_{abc}$, since the factor $1/2$ in (\ref{QCD}) is compensated by
a factor of 2 due to the Majorana nature of the gluino.
Putting the above colour factors together, we obtain
\beq
|{1 \over \sqrt{8}} {f_{had}  \over \sqrt{3}} f_{cda}|^2 = |{1 \over \sqrt{8}} {1 \over \sqrt{3}} 3 \delta_{ch}|^2 = 3 \,. 
\eeq
Note that all color indices are summed over. 

Concerning the spatial part of the wave function, we need to take into account the fact
that both the initial and final states contain two identical particles. In the case of the bound state,
they are in the symmetric $L=0$ state, and the wave function needs to be symmetrised as in
Eq.~(2.14) of~\cite{GoldmanHaber}: 
\beq
{1 \over \sqrt{2}} [\phi_{bs} (r, \theta, \phi) + \phi_{bs} (r, \pi - \theta, \phi + \pi)] = \sqrt{2} \times (\pi a^3)^{-1/2} e^{-r/a}\, . 
\eeq
On the other hand, the final free pair is in the antisymmetric $L=1$ state,
and the wave function needs to be antisymmetrised:
\beq
{1 \over \sqrt{2}} [\phi_f (r, \theta, \phi) - \phi_f (r, \pi - \theta, \phi + \pi)] = \sqrt{2} \times 
 {3 \over 2 |\vec{p}|} R_{p1} (r) P_1 ({\vec{p} \cdot \vec{r} \over |\vec{p}| r}) \, .  
\eeq
The coefficients in these two equations introduce an extra factor of $|\sqrt{2} \sqrt{2}|^2 = 4$
into the modulus-squared of the spatial wave function factors. 
Finally, recall that we have averaged over the polarisations of the gluon, but we also need to average over its colour.
This gives a factor of $1/8$. 
Therefore, the final dissociation cross section is 
\beq
\sigma_{dis} = 3 \times 4 \times {1 \over 8} \times {1 \over 2} \times \sigma^0_{dis} \, ,
\eeq
where the final factor of 1/2 is to avoid double counting of gluinos in the final-state phase-space integration.

\subsection{Formation}

We come finally to the bound-state formation cross section, $\sigma_{bsf}$,
which is related to $\sigma_{dis}$ through the Milne relation: 
\beq
{1\over 2} n_{\tilde g}^{eq} n_{\tilde g}^{eq} \sigma_{bsf} v_{rel} \left(1 + {1 \over e^{\omega/T} - 1} \right) f(v_{rel}) d v_{rel}  = n_{\tilde R}^{eq} \sigma_{dis} d n_g^{eq}  \, ,
\eeq
where the ${1 \over 2}$ on the LHS of the above equation is introduced to avoid double-counting the number of
bound-state formation reactions, and the factor ${1 \over e^{\omega/T} - 1}$ comes from the enhancement of  bound-state
formation due to the stimulated gluon emission in the thermal background (similar to the stimulated recombination in $e^- p \leftrightarrow H \gamma$). 
Using
\bea
d n_g^{eq} & = & g_{g} {4 \pi \over (2 \pi)^3} {\omega^2 d \omega \over e^{\omega/T} - 1}\, , \nonumber \\
f (v_{rel}) & = & \left(m_{\tilde g}/2 \over 2 \pi T \right)^{3/2} 4 \pi v_{rel}^2 e^{-(m_{\tilde g}/2) (v_{rel}^2 / 2 T)}
\eea
and (\ref{conservation}), we find
\beq
\sigma_{bsf} = {2 g_{\tilde{R}} g_{g} \omega^2 \over g^2_{\tilde{g}} [(m_{\tilde g}/2) v_{rel}]^2} \sigma_{dis}  \, , 
\eeq
where
\beq
{g_{\tilde{R}} g_{g} \over g^2_{\tilde{g}}} = {1 \times (2 \times 8) \over (2 \times 8)^2} = {1 \over 16} \, . 
\eeq
For comparison, the Sommerfeld enhanced s-wave cross section for  $\tilde{g}^a \tilde{g}^b \rightarrow g^c g^d$  is
given in Eqs.~(2.13) and (2.25) of~\cite{deSimone:2014pda}:
\beq
S_{ann} (\sigma_{ann} v_{rel}) = \left({1 \over 6} {2 \pi (2 \xi) \over 1 - e^{-2 \pi (2 \xi)}} + {1 \over 3} {2 \pi \xi \over 1 - e^{-2 \pi \xi}} + {1 \over 2} {2 \pi (-{2 \over 3} \xi) \over 1 - e^{2 \pi ({2 \over 3} \xi)}}\right) {(4 \pi \alpha_s)^2 \over m_{\tilde g}^2} {27 \over 512 \pi} \, ,  
\eeq 
where $\xi = {3 \over 2} \alpha_s / v_{rel}$. Therefore, in the $v_{rel} \rightarrow 0$ limit we find
\beq
{\sigma_{bsf} v_{rel} \over S_{ann} (\sigma_{ann} v_{rel})} = \frac{32}{3 e^2} \approx 1.44 \, . 
\eeq
Therefore, we see that the inclusion of $\tilde g \tilde g$ bound states is a non-negligible component in determining the 
final neutralino relic density.

\section{Conversion Rates}

For coannihilation to be effective, the coannihilating species (in this case neutralinos and gluinos)
must be in thermal contact. That is, the rates for interconverting the LSP and NLSP must be faster than
the Hubble rate. In both the familiar cases of stop and stau coannihilation, connectivity of the two species can be taken for granted,
as the conversion rates are mediated by light Standard Model particles and are always fast.
This implies that the ratio of densities ($n_{NLSP}/n_{LSP}$) is approximately equal to
the equilibrium ratio and allows for a simplification in the Boltzmann equations.
However, the interconversion of neutralinos and gluinos must proceed via
squarks, leading to a suppression if the squarks are heavy. The relevance of the coannihilation process relies on fast conversion rates, 
and requires the ratio of squark masses to the gluino mass to be less than approximately 100 as we show below.
For larger squark masses, the gluino and neutralino abundances evolve separately,
and coannihilation effects are essentially shut off independent of the mass difference.

The interconversion processes we consider are $\chi q \leftrightarrow {\tilde g} q$, 
$\chi \bar{q} \leftrightarrow {\tilde g} \bar{q}$, and the gluino decays and the inverse decays
${\tilde g} \leftrightarrow \chi q \bar{q}$. When the neutralino is a Wino or Higgsino, the processes involving a 
chargino, $\chi^+ d \leftrightarrow {\tilde g} u$, $\chi^+ \bar{u}  \leftrightarrow {\tilde g} \bar{d}$ and 
${\tilde g} \leftrightarrow \chi^+ d \bar{u}$, as well as the corresponding processes for $\chi^-$, are also included. 
We note that $q$ stands here for all six quark flavors, and the $u,d$ stand for all the three generations
of up-type and down-type quarks. Also, when $\chi$ is a Higgsino, the two lightest mass-degenerate neutralino components, 
$\tilde{H}_{1,2}$, are both taken into account. 
For each relevant process, we first calculate the
transition matrix element $\tsq$.

We calculate the gluino decay rates for ${\tilde g} \rightarrow \chi q \bar{q}$, ${\tilde g} \rightarrow \chi^+ d \bar{u}$ 
and its charge-conjugated process~\footnote{Since we treat quarks as massless,
three-body decays will always occur as long as $m_{\tilde g} > m_\chi$. Due to logarithmic corrections, 
the gluino two-body decay into a neutralino and a gluon becomes important when the 
squark-to-gluino mass ratio is very large ($\gg$ 100)~\cite{gl2bdecay}. However, as we show below in Section 6,
gluino coannihilation becomes irrelevant for such a large squark-to-gluino mass ratio. 
On the other hand, when the squark-to-gluino mass ratio is $\lesssim 100$, the gluino-neutralino 
conversion rate dominates over the gluino two-body decay rate when the
${\tilde g}$ and $\chi$ are so close in mass that all the gluino three-body decays
would be kinematically forbidden, if quarks are not treated as massless.}. 
The squared transition matrix elements $\tsq$ are identical to the 
corresponding ones for the coannihilation processes given in Appendix B, except that the 
expressions in Appendix B should be multiplied by a factor of 2, because the statistical factor
for the initial spin averaging is $\frac{1}{2} \times \frac{1}{2} = \frac{1}{4}$ for the coannihilation processes, 
whereas it is $\frac{1}{2}$ for the gluino decay processes. We note also
that the definitions of the Mandelstam variables should also be changed correspondingly as follows: 
for the coannihilations, $s = (p_1 + p_2)^2$, $t = (p_1 - p_3)^2$ and $u = (p_1 - p_4)^2$,
whereas for the gluino decays, $s = (p_1 - p_2)^2$, while $t$ and $u$ do not change. 

The gluino decay rates are then obtained by performing the standard 3-body phase space integration. 
The inverse-decay processes do not have to be calculated separately, 
because they are taken into account automatically by the Boltzmann equations given in the next section.  

To calculate the conversion rates for $\chi q \rightarrow {\tilde g} q$, $\chi^+ d \rightarrow {\tilde g} u$, 
$\chi^+ \bar{u} \rightarrow {\tilde g} \bar{d}$ and their charge-conjugated processes, 
we first calculate the cross sections. Again, the squared transition matrix elements $\tsq$ are identical to the 
corresponding ones for the coannihilation processes given in Appendix B, except that the expressions in 
Appendix B should be multiplied by a factor of $\frac{8}{3}$, because the factor for the initial color averaging is 
$\frac{1}{8}$ for the coannihilations, whereas it is $\frac{1}{3}$ for the conversions. 
Also, compared to the coannihilation processes, the Mandelstam variables for the conversion processes 
are re-defined as $s=(p_1 - p_2)^2$, $t=(p_1 \pm p_3)^2$ and $u=(p_1 \mp p_4)^2$, 
where the upper signs in the definition of $t$ and $u$ apply if $\bar{q}_B$ or $\bar{d}_B$ is
brought into the initial state, while the lower signs apply if $q_A$ or $u_A$ is pulled over to the initial state. 

For each of the quark flavors, the thermally-averaged conversion rate is obtained by integrating 
$\sigma_c v_{q}$ over the Fermi-Dirac distribution of the quark in the initial state, 
\beq
\avg{\Gamma_{c}} = \int \sigma_c v_{q} d n_q = \int_{E_{q_{min}}}^{+ \infty} \sigma_c v_{q} \frac{3 \cdot 2 \cdot 4 \pi}{(2 \pi)^3} \frac{|\vec{p}_q|^2 d |\vec{p}_q|}{e^{E_q / T} + 1} \, , 
\label{eq:conversionrate}
\eeq
where $\sigma_c$ is the conversion cross section for any of the relevant processes discussed above. 
In the initial neutralino or chargino rest frame, $\sigma_c$ is a function of the incoming quark energy $E_q$. 
In this reference frame, $t$ or $u$ is the squared center-of-mass energy,
and is given by $m_\chi^2 + m_q^2 + 2 m_\chi E_q$, 
where the lower limit of $E_q$ is $E_{q_{min}} = [(m_{\tilde{g}} + m_{q^{\prime}})^2 - m_\chi^2 - m_q^2] /(2 m_\chi)$, 
where $q^{\prime}$ represents the quark in the final state. Here $v_q$ is the velocity of the incoming quark, 
and it is related to the energy and 3-momentum of the quark by $v_q = |\vec{p}_q| / E_q$. 
The factors 3 and 2 in (\ref{eq:conversionrate}) count the quark color and spin degrees of freedom, respectively. 
Again, the inverse conversion rates are taken into account automatically by the Boltzmann equations.

\section{Boltzmann Equations}

We are now in a position to put all of the components discussed above into a rate equation (or set of equations)
in order to solve for the relic density.
To do so, we begin by considering three separate density components: neutralinos, gluinos and bound states. 

To set up a coupled set of Boltzmann equations, it is convenient to rescale the number densities of 
neutralinos, gluinos and bound states by the entropy density,
\beq
Y_\chi \equiv {n_\chi \over s}, \; Y_{\tilde g} \equiv {n_{\tilde g} \over s}, \; Y_{\tilde R} \equiv {n_{\tilde R} \over s} \, .
\label{yields}
\eeq
These are governed by the following set of coupled Boltzmann equations~\footnote{A set of coupled Boltzmann equations for the photino and a gluino $R$-hadron was studied in~\cite{cfk}. In a different context, Boltzmann equations involving a bound state can be found, for example, in~\cite{dmphp}.}:
\bea
{d Y_\chi \over d x} &=& 
{x s \over H(m_\chi)}
\Big[
 - {\avg {\sigma v}}_{\chi \chi} \left( Y_\chi Y_\chi - Y_\chi^{eq} Y_\chi^{eq} \right)  
 - {\avg {\sigma v}}_{\chi {\tilde g}} \left( Y_\chi Y_{\tilde g}- Y_\chi^{eq} Y_{\tilde g}^{eq} \right)  
 \nonumber  \\ &&\quad\quad\quad\quad
 - \sum_{q}\avg{\Gamma_{c}} {1 \over s} \left( Y_\chi - Y_\chi^{eq} {Y_{\tilde g} \over Y_{\tilde g}^{eq}} \right)  
 + {\avg \Gamma}_{\tilde g} {1 \over s} \left( Y_{\tilde g} - Y_{\tilde g}^{eq} { Y_\chi \over Y_\chi^{eq}} \right) 
 \Big] \, , 
\nonumber \\ 
\label{eq:dchidx} 
 \\
{d Y_{\tilde g} \over d x} &=& 
{x s \over H(m_\chi)}
\Big[
 - {\avg {\sigma v}}_{{\tilde g} {\tilde g}} \left( Y_{\tilde g} Y_{\tilde g} - Y_{\tilde g}^{eq} Y_{\tilde g}^{eq} \right)  
 - {\avg {\sigma v}}_{\chi {\tilde g}} \left( Y_\chi Y_{\tilde g}- Y_\chi^{eq} Y_{\tilde g}^{eq} \right)  
 \nonumber  \\ &&\quad\quad\quad\quad
 +  \sum_{q}\avg{\Gamma_{c}} {1 \over s} \left( Y_\chi - Y_\chi^{eq} {Y_{\tilde g} \over Y_{\tilde g}^{eq}} \right)  
 - {\avg \Gamma}_{\tilde g} {1 \over s} \left( Y_{\tilde g} - Y_{\tilde g}^{eq} { Y_\chi \over Y_\chi^{eq}} \right) 
 \nonumber  \\ &&\quad\quad\quad\quad
 - {\avg {\sigma v}}_{bsf} \left( Y_{\tilde g} Y_{\tilde g} - Y_{\tilde g}^{eq} Y_{\tilde g}^{eq} {Y_{\tilde R} \over Y_{\tilde R}^{eq}} \right)  
 \Big] \, ,  
\label{eq:dgldx} 
 \\
 {d Y_{\tilde R} \over d x} &=& 
{x s \over H(m_\chi)}
\Big[ 
 - {\avg \Gamma}_{\tilde R} {1 \over s} \left( Y_{\tilde R} - Y_{\tilde R}^{eq} \right) 
 - {\avg {\sigma v}}_{{\tilde g} {\tilde R} \rightarrow {\tilde g} g} Y_{\tilde g} \left( Y_{\tilde R}  - Y_{\tilde R}^{eq} \right)  
 \nonumber  \\ &&\quad\quad\quad\quad
+ {1 \over 2} {\avg {\sigma v}}_{bsf} \left( Y_{\tilde g} Y_{\tilde g} - Y_{\tilde g}^{eq} Y_{\tilde g}^{eq} {Y_{\tilde R} \over Y_{\tilde R}^{eq}} \right) 
 \Big] \, ,  
\label{eq:dbsdx} 
\eea
where
\beq
x \equiv {m_\chi \over T}, \, s = {2 \pi^2 \over 45} g_{\ast s} T^3, \; H(m_\chi) \equiv H(T) x^2 = \left({4 \pi^3 G_N g_\ast \over 45}\right)^{1 \over 2} m_\chi^2 \, , 
\eeq
and $g_{\ast s}$ and $g_\ast$ are the total numbers of effectively massless degrees of freedom associated with
the entropy density and the energy density, respectively, 
${\avg {\sigma v}}_{\chi \chi}$ is the relative velocity times the total cross section for the channels
for $\chi \chi$ annihilation into Standard Model particles, and
${\avg {\sigma v}}_{\chi {\tilde g}}$ and ${\avg {\sigma v}}_{{\tilde g} {\tilde g}}$ are to be understood similarly, 
$\sum_{q} \avg{\Gamma_{c}}$ and ${\avg \Gamma}_{\tilde g}$ are the total conversion rate and gluino decay rate discussed in the previous section, and all possible quark and anti-quark channels for the $\chi$ are summed over, 
${\avg \Gamma}_{\tilde R}$ is the decay rate of the $\tilde R$, and   
${\avg {\sigma v}}_{bsf}$ is the bound-state formation cross section times the relative velocity of the two incoming gluinos, taking into account the $1/(e^{\omega / T} -1)$ enhancement factor as discussed in Section 3.2.   
Finally, ${\avg {\sigma v}}_{{\tilde g} {\tilde R} \rightarrow {\tilde g} g} Y_{\tilde g}$ has the same effect as ${\avg \Gamma}_{\tilde R} /s$, namely, it converts the bound states to gluons without altering the density of free gluinos. All the quantities bracketed by $\langle \dots \rangle$ 
are thermally averaged, and the superscript `eq' denotes equilibrium yields. 

Eq. (\ref{eq:dbsdx}) can be written in a more intuitive form:
\beq
 {d \ln Y_{\tilde R} \over d \ln x} = 
-{{\avg \Gamma}_{\tilde R} + {\avg {\sigma v}}_{{\tilde g} {\tilde R} \rightarrow {\tilde g} g} n_{\tilde g} \over H(T)} 
\left( 1 - {Y_{\tilde R}^{eq} \over Y_{\tilde R}} \right)
+ \frac{{1 \over 2} {\avg {\sigma v}}_{bsf} n_{\tilde g} \left(Y_{\tilde g} \over Y_{\tilde R} \right)}{H(T)} 
\left[ 1 - \left(Y_{\tilde g}^{eq} \over Y_{\tilde g} \right)^2 \left(Y_{\tilde R} \over Y_{\tilde R}^{eq} \right)   \right] \, . 
\label{eq:dlnbsdlnx}
\eeq
One can check that the LHS of Eq.~(\ref{eq:dlnbsdlnx}) is of order -10, whereas each of the terms on the RHS of Eq.~(\ref{eq:dlnbsdlnx}) are of order $\alpha_s^5 M_P/m_\chi$,
where $M_P = G_N^{-1/2}$.
Hence, to a good approximation, we can set the two terms on the RHS equal to each other and solve for
$\left(Y_{\tilde R} \over Y_{\tilde R}^{eq} \right)$:
\beq
{Y_{\tilde R} \over Y_{\tilde R}^{eq}} = {{{\avg \Gamma}_{\tilde R} + {\avg {\sigma v}}_{{\tilde g} {\tilde R} \rightarrow {\tilde g} g} n_{\tilde g} + {\avg \Gamma}_{dis}} \left(Y_{\tilde g} \over Y_{\tilde g}^{eq} \right)^2 
\over {\avg \Gamma}_{\tilde R} + {\avg {\sigma v}}_{{\tilde g} {\tilde R} \rightarrow {\tilde g} g} n_{\tilde g} + {\avg \Gamma}_{dis}} \, ,  
\eeq
where 
\beq
{\avg \Gamma}_{dis} = {1 \over 2} {\avg {\sigma v}}_{bsf} (n_{\tilde g}^{eq})^2 / n_{\tilde R}^{eq} \, . 
\eeq
Therefore, we find that
\bea
 {d (Y_\chi + Y_{\tilde g}) \over dx} &=& {x s \over H(m_\chi)} 
 \Big[ -\sum_{i,j = \chi, \tilde g} {\avg {\sigma v}}_{ij} \left(Y_i Y_j - Y_i^{eq} Y_j^{eq} \right) 
  \nonumber  \\ &&\quad\quad\quad\quad
 - {\avg {\sigma v}}_{bsf} {{\avg \Gamma}_{\tilde R} + {\avg {\sigma v}}_{{\tilde g} {\tilde R} \rightarrow {\tilde g} g} n_{\tilde g}  \over {\avg \Gamma}_{\tilde R} + {\avg {\sigma v}}_{{\tilde g} {\tilde R} \rightarrow {\tilde g} g} n_{\tilde g} + {\avg \Gamma}_{dis}} \left( Y_{\tilde g} Y_{\tilde g} - Y_{\tilde g}^{eq} Y_{\tilde g}^{eq} \right)
\Big] \, . 
\label{chig}
\eea
Moreover, we note that ${\avg {\sigma v}}_{{\tilde g} {\tilde R} \rightarrow {\tilde g} g} n_{\tilde g}$ is much
smaller than ${\avg \Gamma}_{\tilde R}$ for $x \gtrsim 10$, due to the fact that $n_{\tilde g}$ 
decreases with the decrease of temperature while ${\avg \Gamma}_{\tilde R}$ is nearly temperature independent. Since the process ${\tilde g} {\tilde R} \rightarrow {\tilde g} g$ is related to the bound-state formation process by crossing, ${\avg {\sigma v}}_{{\tilde g} {\tilde R} \rightarrow {\tilde g} g}$ should be related to ${\avg {\sigma v}}_{bsf}$ by a coefficient not too much different from order 1. 

If at least one of the $\sum_{q} \avg{\Gamma_{c}}$ and ${\avg \Gamma}_{\tilde g}$
is sufficiently larger than $H(T)$ throughout the period during which ($Y_\chi + Y_{\tilde g})$ changes substantially, which is the case when the squark mass appearing
in the denominators of the matrix elements for these processes is not too large, Eq.~(\ref{chig}) can be solved by using the very good approximation ${Y_{\tilde g}}/{Y_\chi} \approx {Y_{\tilde g}^{eq}}/{Y_\chi^{eq}}$. In this case, Eq.~(\ref{chig}) can be recast in the familiar form suitable for 
coannihilation calculations, and we can write 
\beq
{d Y \over dx} = - {x s \over H(m_\chi)} \left(1- \frac{x}{3 g_{\ast s}} \frac{d g_{\ast s}}{d x} \right) \langle\sigma_{\rm eff} v_{\rm rel}\rangle \left(Y^2 - (Y_{eq})^2 \right) \, ,
\label{rate2}
\eeq
where we have included the term $\frac{x}{3 g_{\ast s}} \frac{d g_{\ast s}}{d x}$ which takes into account the evolution of $g_{\ast s}$ with temperature. As we will see, this approximation is valid so long as
$m_{\tilde q}/m_{\tilde g} \la 20$. 

In Eq.~(\ref{rate2}), $Y = n/s$, where $n$ is interpreted as the total number density,
\beq
n \equiv \sum_i n_i = n_\chi + n_{\tilde g} \, .
\label{n}
\eeq
and $Y_{eq} = \neq/s$, where $\neq$ is the total equilibrium number density,
\beq
\neq \equiv  \sum_i n_{{\rm eq},i} = n_\chi^{eq} + n_{\tilde g}^{eq} \, .
\label{neq}
\eeq
The effective annihilation cross section is
\beq
\langle\sigma_{\rm eff} v_{\rm rel}\rangle \equiv
\sum_{ij}{ n_{{\rm eq},i} n_{{\rm eq},j} \over \neq^2}
\langle\sigma_{ij} v_{\rm rel}\rangle \, .
\label{sv2} 
\eeq
As one can see from Eq.~(\ref{chig}), the expression for $\langle \sigma v \rangle_{\tilde g \tilde g}$ 
is the `standard' term in the first line of (\ref{chig}) combined with the second line involving the bound states.
We re-emphasise that this simplification requires a fast interconversion rate as discussed in the previous Section, so that
we can set $(Y_{\tilde g}/Y_\chi) = (Y^{eq}_{\tilde g}/Y^{eq}_\chi)$, which is true only when $m_{\tilde q}/m_{\tilde g} \la 20$.
For larger squark masses, we use the coupled set of Boltzmann equations to solve for the relic density. 

When the LSP is a Wino or a Higgsino, we can still use all the above equations to solve for the relic density. All we need to do is re-define the following quantities to include the contributions from each of the $\chi$ components, $\chi_i$, neutral or charged, as
\bea
n_\chi & \equiv & \sum_{\chi_i} n_{\chi_i} \, , \nonumber \\
n_\chi^{eq} & \equiv & \sum_{\chi_i} n_{\chi_i}^{eq} \, , \nonumber \\
{\avg {\sigma v}}_{\chi \chi} & \equiv & \sum_{\chi_i, \chi_j} {\avg {\sigma v}}_{\chi_i \chi_j} r_{\chi_i} r_{\chi_j} \, \nonumber \\
{\avg {\sigma v}}_{\chi {\tilde g}} & \equiv & \sum_{\chi_i} {\avg {\sigma v}}_{\chi_i {\tilde g}} r_{\chi_i} \, , \nonumber \\
\avg{\Gamma_{c}} & \equiv & \sum_{\chi_i} \avg{\Gamma_{c}}_{\chi_i q \to {\tilde g} q^\prime} r_{\chi_i} \, , \nonumber \\
{\avg \Gamma}_{\tilde g} & \equiv & \sum_{\chi_i} {\avg \Gamma}_{{\tilde g} \rightarrow \chi_i q \bar{q}^\prime} \, ,
\label{eq:multicomponents}
\eea
where $q^\prime$ is the same as $q$ when $\chi_i$ is a neutralino, and they are different when $\chi_i$ is a chargino, and the $q$ and $q^\prime$ indicate all the possible quark and anti-quark channels for the conversion rates and gluino decay rates. 
In Eq.~(\ref{eq:multicomponents}), $r_{\chi_i} \equiv n_{\chi_i}^{eq} / n_\chi^{eq} = n_{\chi_i} / n_\chi$, where the latter `$=$' is guaranteed by the fast conversion and/or decay rates among the different $\chi_i$'s. For later discussion, it is useful to define an effective number of degrees of freedom for $\chi$:
\beq
g_{\chi_{eff}} \equiv \sum_{\chi_i} g_{\chi_i} \left(1+ \Delta_{\chi_i}\right)^{3/2} \exp(-\Delta_{\chi_i} m_{\chi_1}/T) \, ,
\label{eq:gchieff} 
\eeq
where $\Delta_{\chi_i} \equiv (m_{\chi_i} / m_{\chi_1} - 1)$, and we assume $\chi_1$ is the lightest component (i.e., the LSP) among the $\chi_i$'s. We can then write $r_{\chi_i}$ explicitly as 
\beq
r_{\chi_i} = \frac{g_{\chi_i}}{g_{\chi_{eff}}} \left(1+ \Delta_{\chi_i}\right)^{3/2} \exp(-\Delta_{\chi_i} m_{\chi_1}/T) \, .
\eeq
In the limit that all the $\chi$ components have the same mass, $g_{\chi_{eff}} =2$, 6 and 8 for Bino, Wino and Higgsino, respectively.


\section{Numerical Results}

We now present some numerical results obtained using the above formalism.
Our results in this section are based on simplified supersymmetric spectra
defined at the weak scale. We assume degenerate squark masses, $m_{\tilde q}$
and for the most part, our results do not depend on supersymmetric parameters such as
$\mu$, $A_0$, and $\tan \beta$\footnote{The exception is the case of Higgsino-gluino
coannihilations for which the vertices do depend on $\tan \beta$.}. We assume that the 
neutralino is a pure state of either a Bino, Wino, or Higgsino. Thus our free parameters
are simply the neutralino mass, $m_\chi$, the gluino mass, $m_{\tilde g}$ and the squark masses,
$m_{\tilde q}$. In future work, we apply these results to more realistic CMSSM-like models 
(without gaugino mass universality) and pure gravity mediation models with vector-like multiplets
\cite{eelo}.  

 We begin with
the case in which the lightest neutralino $\chi$ is the Bino. 

\subsection{Bino LSP}

\begin{figure}
\begin{center}
\begin{tabular}{c c}
\hspace{-0.6cm}
\includegraphics[height=5cm]{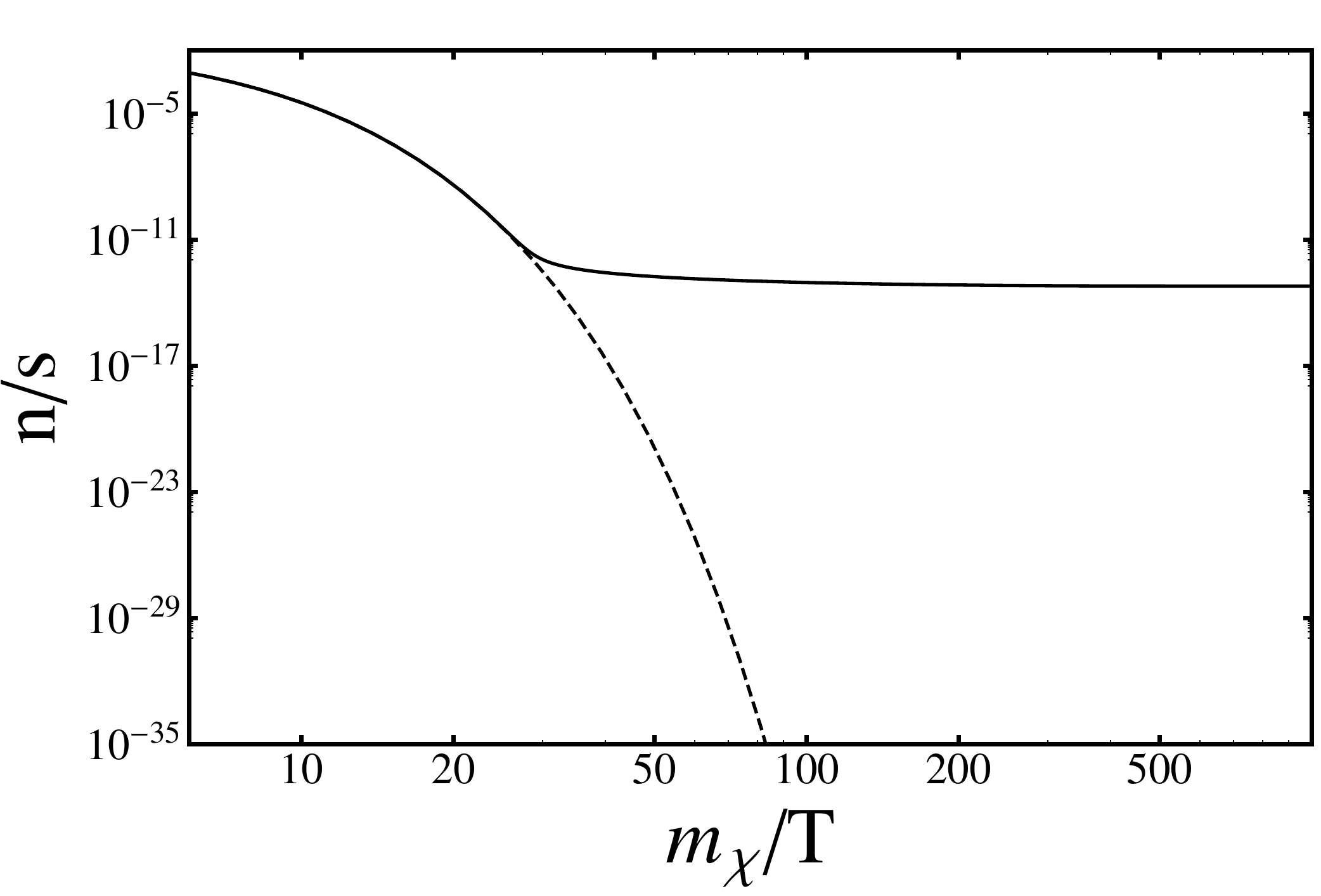} & 
\hspace{-0.6cm}
\includegraphics[height=5cm]{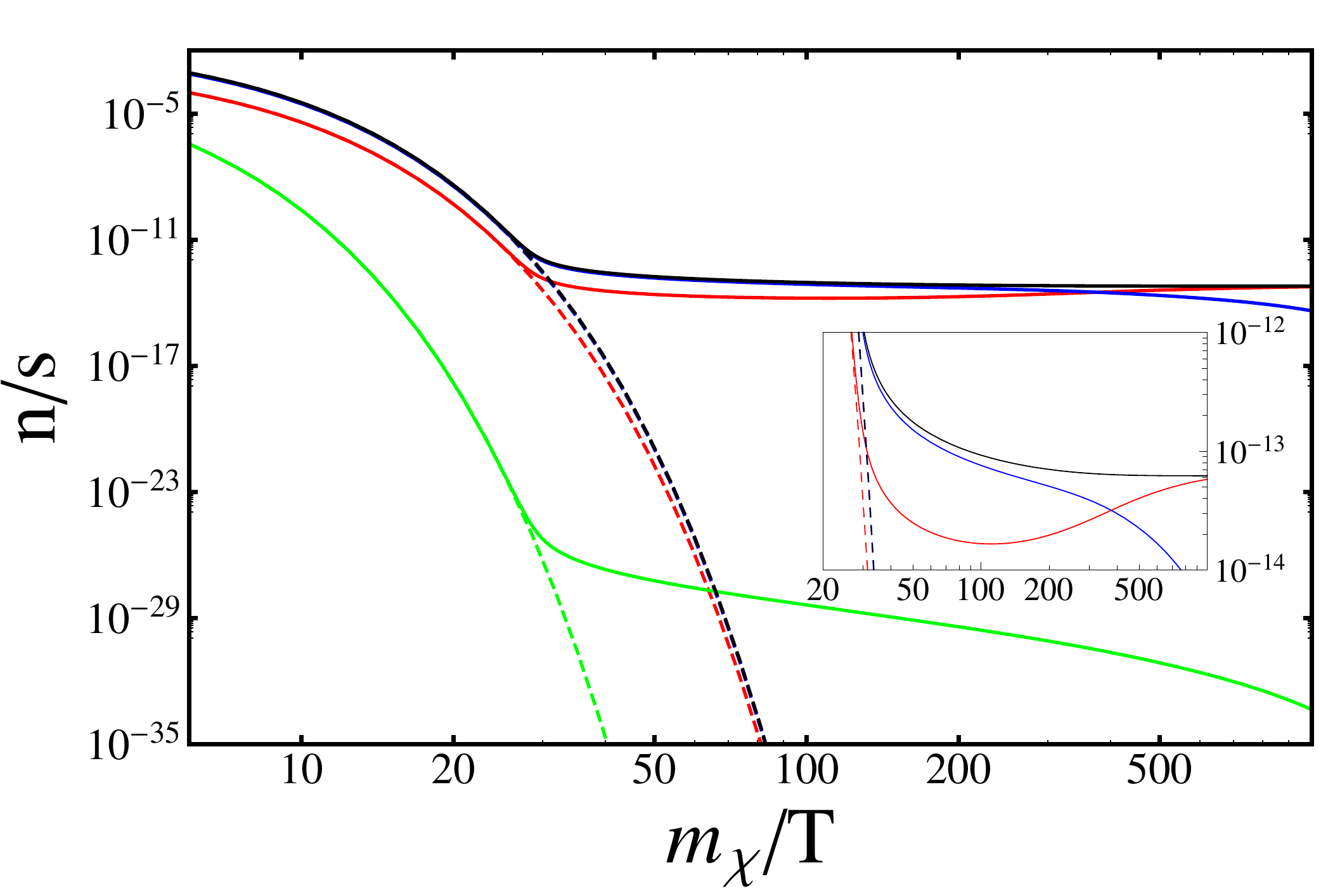} \\
\end{tabular}
\end{center}   
\caption{\label{fig:singletriple}\it
Left panel: The evolution of the total supersymmetric particle abundance $n/s$ as a function
of $m_\chi/T$ for the representative case $m_\chi = 7$~TeV, $\Delta m \equiv m_{\tilde g} - m_\chi = 40$~GeV,
and $m_{\tilde q}/m_{\tilde g} = 10$
using the single Boltzmann equation (\ref{chig}). The dashed line exhibits the naive thermal equilibrium abundance,
and the solid line shows the numerical solution  of equation (\ref{chig}) for the sum of the 
neutralino and gluino densities, exhibiting the familiar freeze-out when $m/T \sim 30$. 
Right panel: the separate evolutions of the 
abundances (solid lines) using the full set of equations given by Eqs. (\ref{eq:dchidx} - \ref{eq:dbsdx}) for 
the gluino (blue lines), Bino (red lines), gluino-gluino bound-state (green lines) and sum of the Bino and 
gluino densities (black lines) as functions
of $m_\chi/T$. Details of the evolutions of the abundances are shown in the inset.
The dashed lines again exhibit the naive thermal equilibrium abundances.
}
\end{figure}

Fig.~\ref{fig:singletriple} compares a naive calculation using a single
Boltzmann equation for the total relic abundance (left panel) with a treatment of the three
coupled Boltzmann equations for the gluino, Bino and gluino-gluino bound state
abundances (right panel). These results are for the representative case $m_\chi = 7 \tev$,
$\Delta m \equiv m_{\tilde g} - m_\chi = 40 \gev$ and $m_{\tilde q}/m_{\tilde g} =10$.
The dashed line in the left panel shows the total relic abundance
as would be given by the Boltzmann distribution if thermal equilibrium were maintained, and we see
a clear departure for $m/T \gtrsim 30$, as expected from a freeze-out calculation.
In the right panel of Fig.~\ref{fig:singletriple}, the blue line shows the evolution of the gluino abundance,
the red line that of the Bino abundance,  the green line that of the
gluino-gluino bound states, and the black line that of the sum of the gluino and Bino densities.
The inset shows details of the evolutions of the gluon, Bino and bound-state abundances.
The dashed lines again show the corresponding naive thermal equilibrium abundances.
With the stated choices of $m_{\tilde g}$, $m_\chi$ and $m_{\tilde q}$, the relic dark matter
density using the single Boltzmann equation is $\Omega_\chi h^2 = 0.120$, and the
full set of three Boltzmann equations yields $\Omega_\chi h^2 = 0.119$.
Correspondingly, the black lines in the right panel of Fig.~\ref{fig:singletriple} are
indistinguishable from the lines in the left panel.

\begin{figure}
\begin{center}
\begin{tabular}{c c}
\hspace{-0.6cm}
\includegraphics[height=5cm]{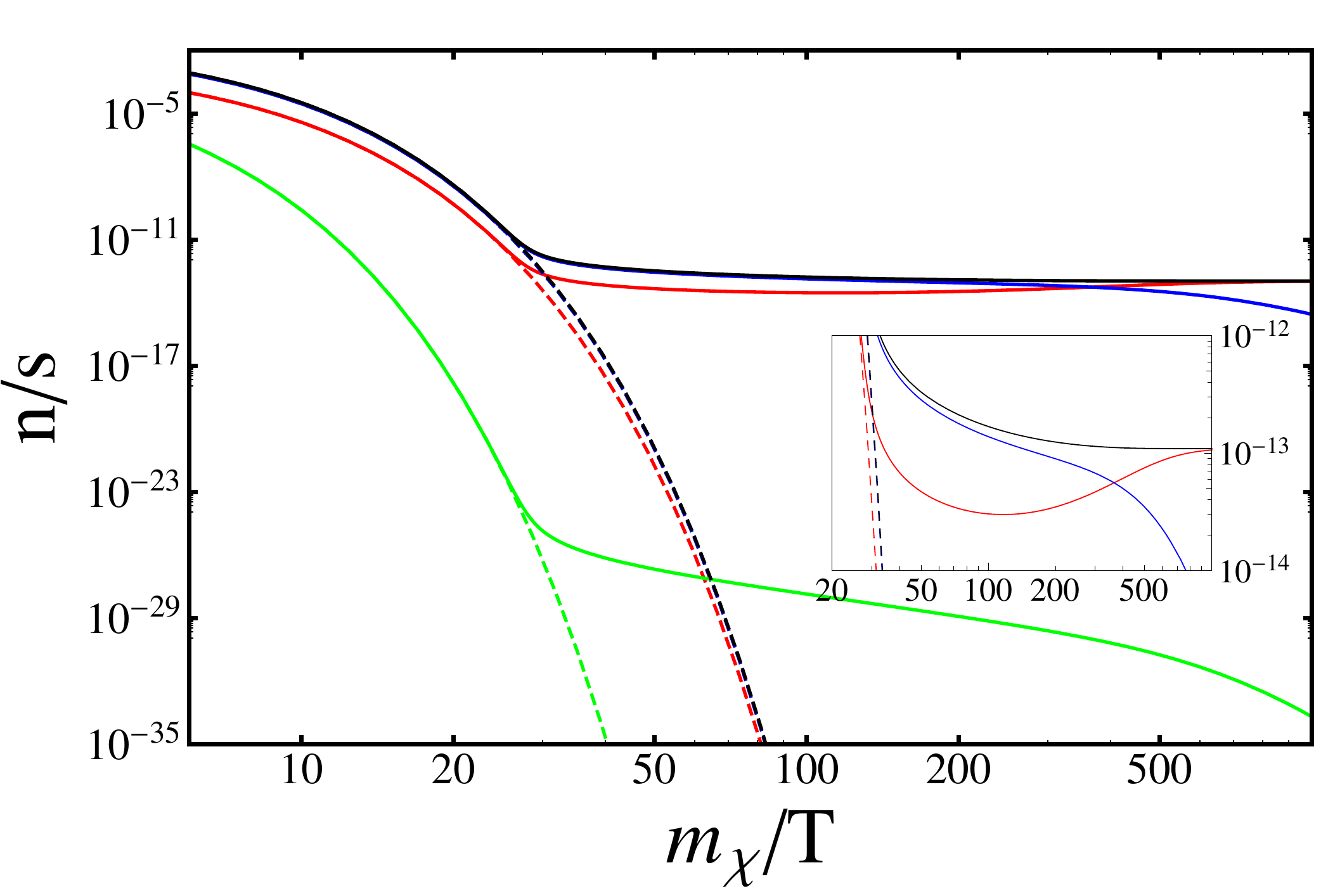} & 
\hspace{-0.6cm}
\includegraphics[height=5cm]{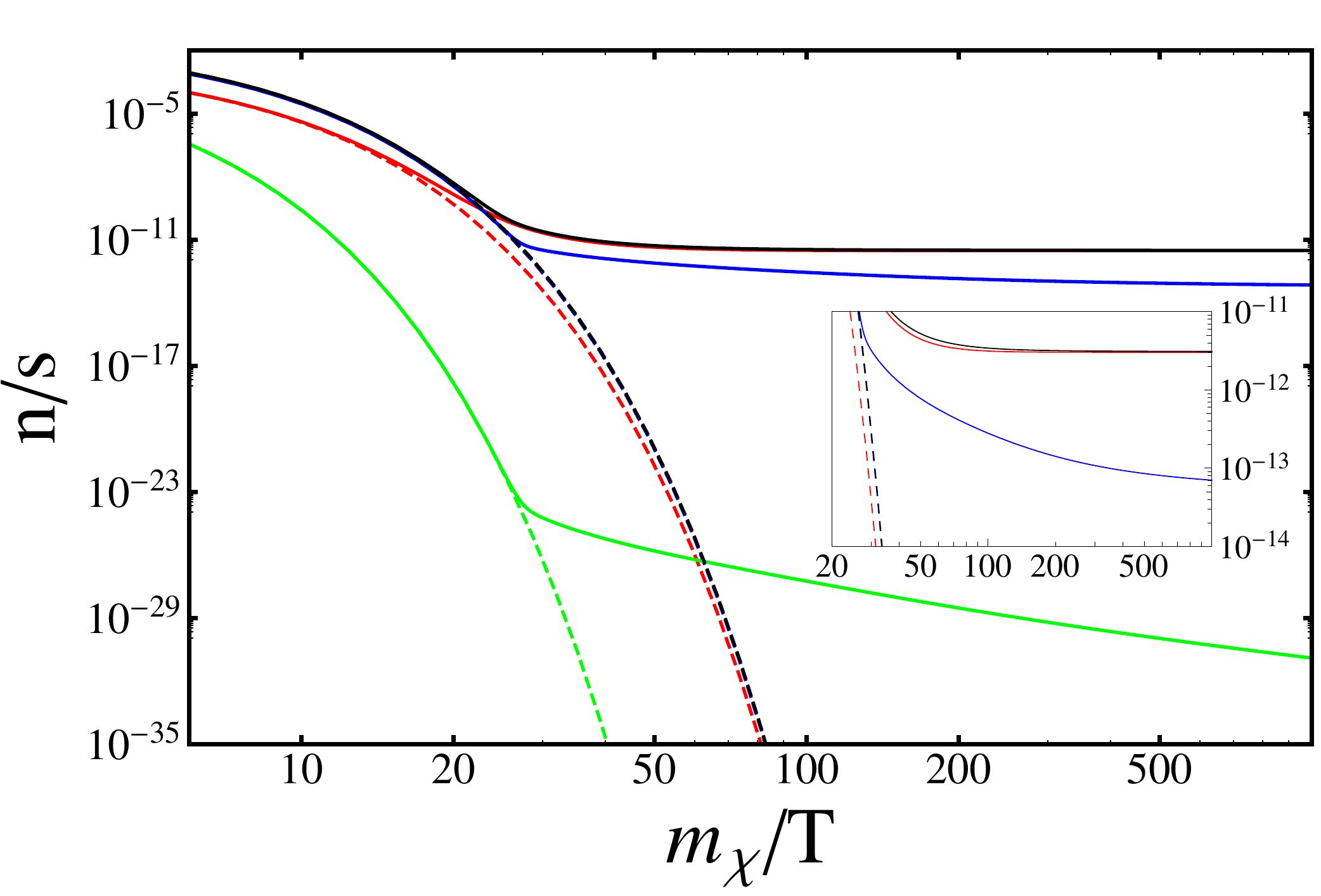} \\
\end{tabular}
\end{center}   
\caption{\label{fig:varymsq}\it
As in the right panel of Fig.~\protect\ref{fig:singletriple}, but for the choices
$m_{\tilde q}/m_{\tilde g} = 1.1$ (left), $120$ (right).
}
\end{figure}

Fig.~\ref{fig:varymsq} shows the effect of varying $m_{\tilde q}/m_{\tilde g}$ for the same
representative values $m_\chi = 7 \tev$ and $\Delta m \equiv m_{\tilde g} - m_\chi = 40 \gev$.
The left panel is for $m_{\tilde q}/m_{\tilde g} = 1.1$~\footnote{We do not show
results for smaller values of $m_{\tilde q}/m_{\tilde g}$, since then squark-Bino coannihilations
should also be taken into account.}, in which case we find that the
relic cold dark matter density is higher than previously:  $\Omega_\chi h^2 = 0.21$.
This is due to the fact that at a low squark to gluino mass ratio, there is a cancellation among
the $t$ and $u$ channel annihilations with the $s$ channel leading to a smaller gluino annihilation
cross section and hence a larger relic density. 
The results also change, even more significantly, for large values of $m_{\tilde q}/m_{\tilde g}$,
as shown in the right panel of Fig.~\ref{fig:varymsq}, where $m_{\tilde q}/m_{\tilde g} = 120$.
In this case, we find a much larger value of $\Omega_\chi h^2 = 6.0$.
In both panels, the insets show details of the evolutions of the gluino, Bino and bound-state abundances.
At still larger values of $m_{\tilde q}/m_{\tilde g}$ the relic density grows very sharply: for example, for
$m_{\tilde q}/m_{\tilde g} = 150$, we find $\Omega_\chi h^2 =  780$, and
for $m_{\tilde q}/m_{\tilde g} = 200$  we find $\Omega_\chi h^2 = 1.0 \times 10^5$
assuming the same gluino-Bino mass difference of 40 GeV.
These large numbers reflect the failure of gluino-neutralino conversion to keep pace
with the Hubble expansion for large $m_{\tilde q}/m_{\tilde g}$.

\begin{figure}
\begin{center}
\includegraphics[height=7cm]{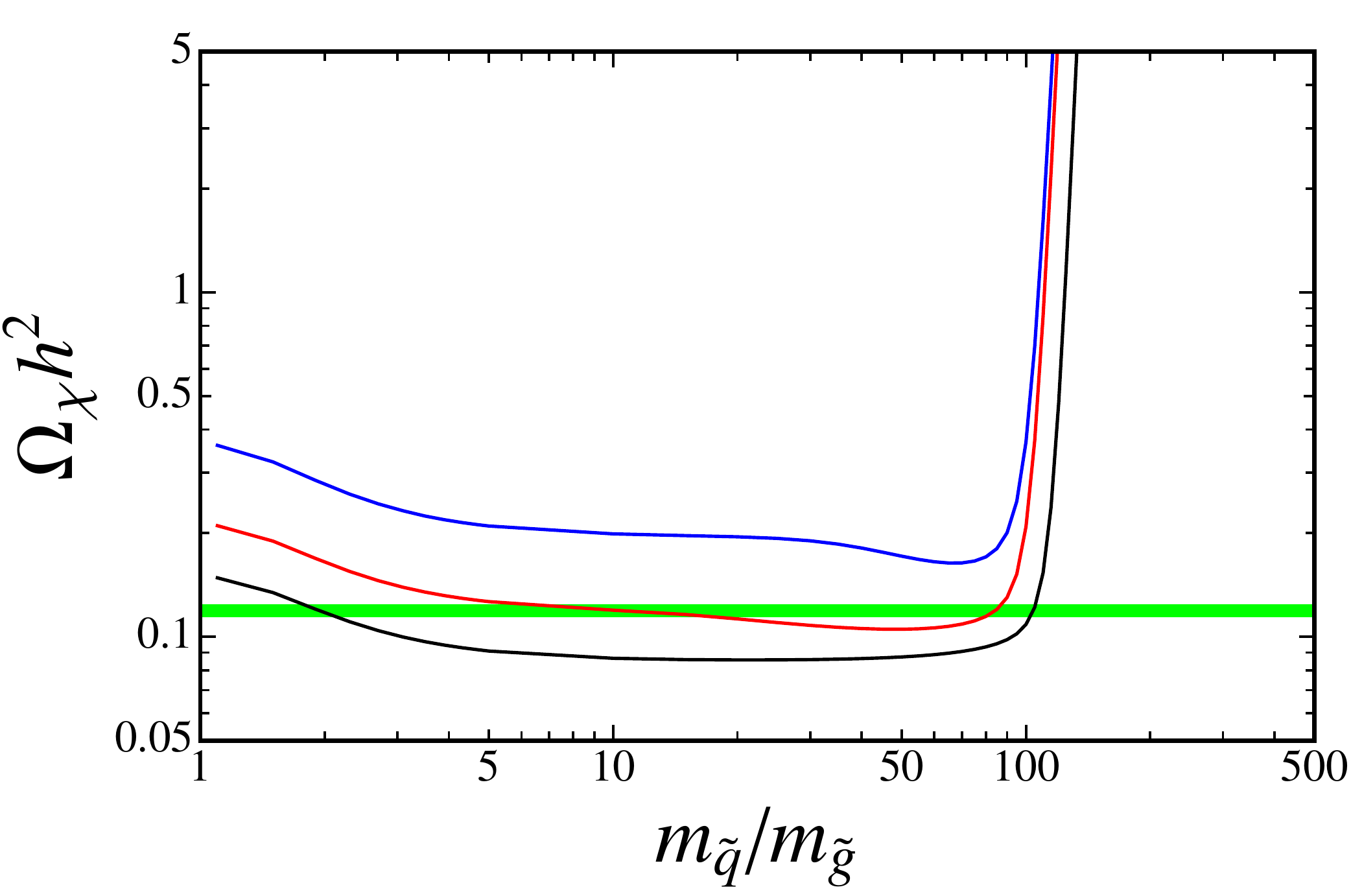}
\end{center}   
\caption{\label{fig:basin}\it
The relic cold dark matter density $\Omega_\chi h^2$ as a function of $m_{\tilde q}/m_{\tilde g}$ for $m_\chi = 7$~TeV
and the choics $\Delta m \equiv m_{\tilde g} - m_\chi = 0, 40$, and 120~GeV (from bottom to top, black, red,
and blue lines, respectively).
The rise at small $m_{\tilde q}/m_{\tilde g}$ is due to the cancellations between the $s$-, $t$-
and $u$-channel diagrams for gluino pair annihilation, and the rise at large $m_{\tilde q}/m_{\tilde g}$ is due to the
decoupling of the gluino and neutralino densities.
The horizontal band indicates the 3-$\sigma$ range for the Planck determination of the 
cold dark matter density of $\Omega h^2 = 0.1193 \pm 0.0014$ \cite{planck}. 
}
\end{figure}

In order to summarize the effects of both the cancellations in the annihilation cross section at low $m_{\tilde q}/m_{\tilde g}$
and the decoupling of the gluino coannihilations at high $m_{\tilde q}/m_{\tilde g}$, we show in Fig.~\ref{fig:basin},
the relic neutralino density as a function of $m_{\tilde q}/m_{\tilde g}$ for our nominal value of 
$m_\chi = 7 \tev$, and $\Delta m \equiv m_{\tilde g} - m_\chi = 0, 40$, and 120  $\gev$
(black, red, and blue lines, respectively). We see clearly the rise in $\Omega_\chi h^2$ at small
$m_{\tilde q}/m_{\tilde g}$ as well as the very rapid rise in $\Omega_\chi h^2$ at high $m_{\tilde q}/m_{\tilde g} \gtrsim 100$.
In between there is a plateau with lower $\Omega_\chi h^2$, as exemplified by the case $m_{\tilde q}/m_{\tilde g} = 10$
shown in Fig.~\ref{fig:singletriple}. In general, there is a shallow minimum in $\Omega_\chi h^2$ around $m_{\tilde q}/m_{\tilde g} \sim 50$
whose location depends on $\Delta m$.
The horizontal band indicates the 3-$\sigma$ range for the Planck determination of the 
cold dark matter density of $\Omega h^2 = 0.1193 \pm 0.0014$ \cite{planck}. 

\begin{figure}
\begin{center}
\begin{tabular}{c c}
\hspace{-0.6cm}
\includegraphics[height=8cm]{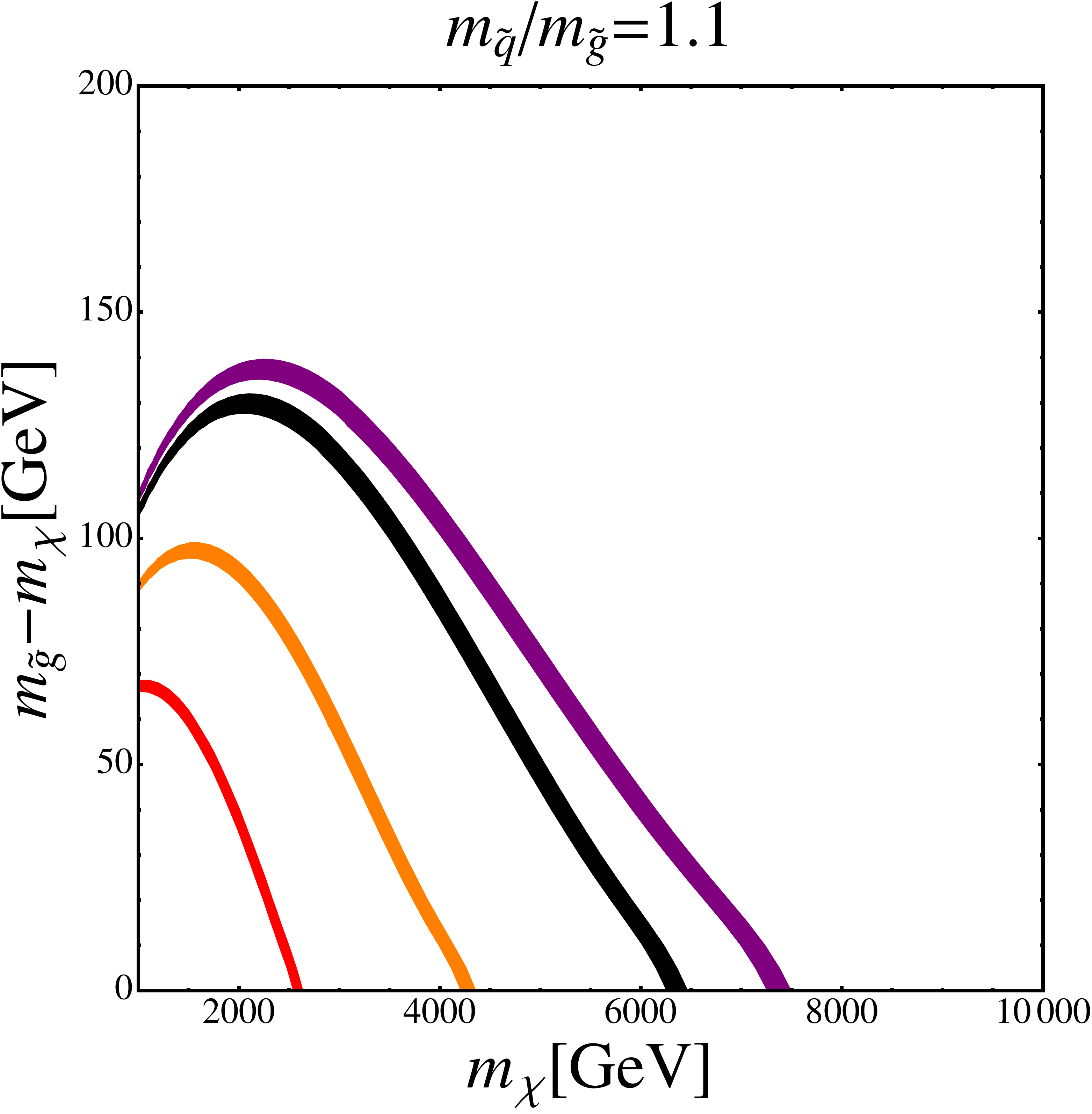} & 
\hspace{-0.6cm}
\includegraphics[height=8cm]{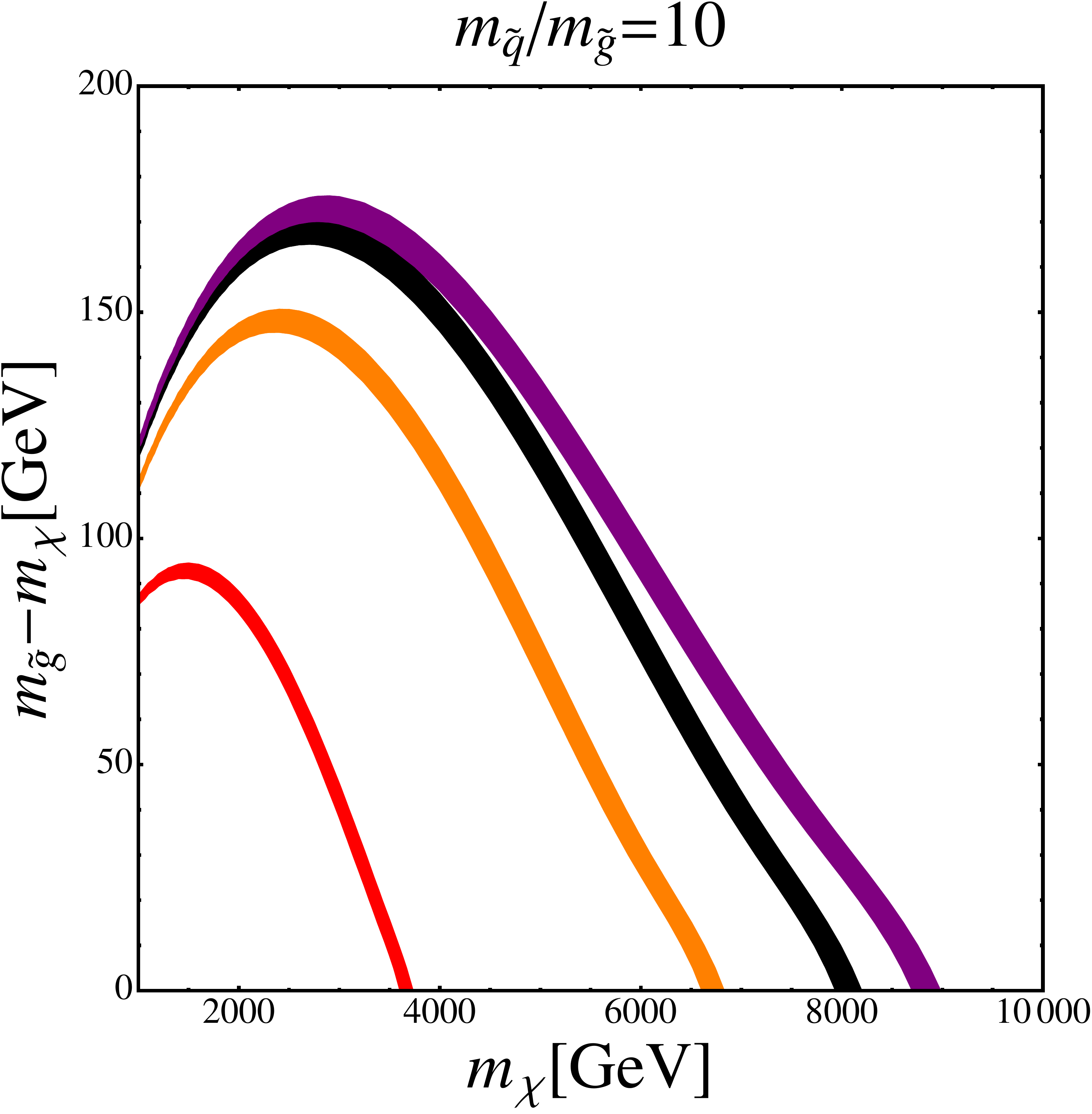} \\
\hspace{-0.6cm}
\includegraphics[height=8cm]{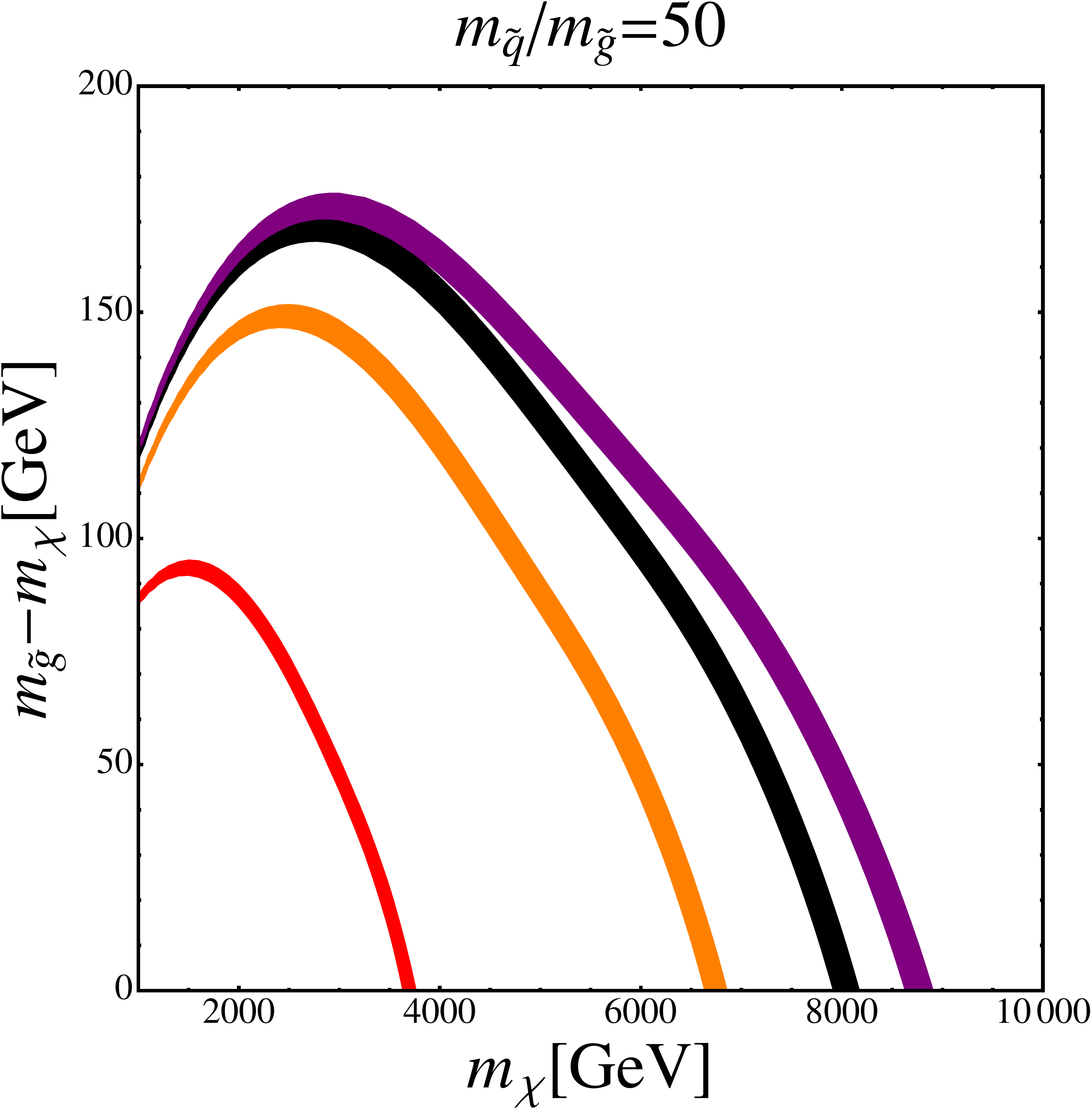} &
\hspace{-0.6cm}
\includegraphics[height=8cm]{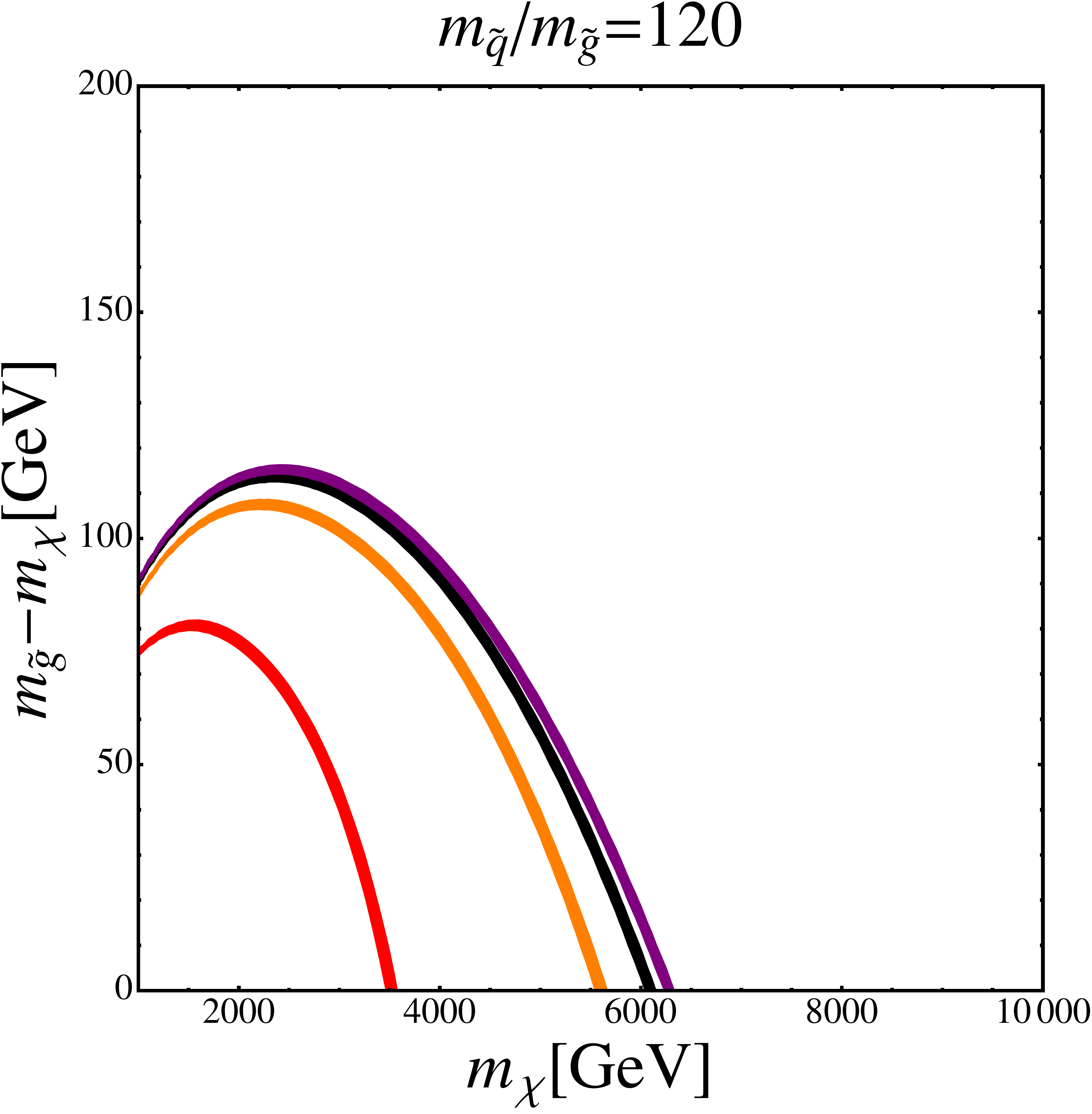} 
\end{tabular}
\end{center}   
\caption{\label{fig:bino}\it
The $(m_\chi, \Delta m \equiv m_{\tilde g} - m_\chi)$ planes
for a Bino LSP, exhibiting bands where $0.1151 < \ohsq < 0.1235$ (3 $\sigma$ below and above the current central value),
for different values of $m_{\tilde q}/m_{\tilde g} = 1.1$
(upper left), $10$ (upper right), $50$ (lower left) and $120$ (lower right). These results are
calculated without the Sommerfeld enhancement factor and gluino bound-state formation (red bands),
with the Sommerfeld enhancement factor but without gluino bound-state formation (orange bands),
with both the Sommerfeld enhancement factor and gluino bound-state formation (black bands),
and allowing for the possibility that the bound-state formation rate is a factor 2 larger than our calculations (purple bands).
}
\end{figure}

The panels of Fig.~\ref{fig:bino} display bands in the $(m_\chi, \Delta m)$ plane
where $0.1151 < \ohsq < 0.1235$ (3 $\sigma$ below and above the current central value \cite{planck})
for a selection of values of $m_{\tilde q}/m_{\tilde g}$, as
calculated in various dynamical approximations. The red bands were calculated dropping both
the Sommerfeld enhancement factor and the effect of gluino bound-state formation. As was already
noted in~\cite{deSimone:2014pda,EOZ} the Sommerfeld enhancement causes a significant
suppression of $\ohsq$ for fixed values of the model parameters, so long as the gluino-Bino conversion
rates are large enough that the gluino coannihilation is effective. Correspondingly, the
orange $\ohsq$ bands, calculated including the Sommerfeld factor,
appear at larger values of $\Delta m$ and extend to larger values of $m_\chi$. The
effect of including bound-state effects is to suppress further the value of $\ohsq$
for fixed model parameters, so that the
corresponding black $\ohsq$ bands in Fig.~\ref{fig:bino} extend to even larger values of $\Delta m$
and $m_\chi$. We also show in Fig.~\ref{fig:bino}
(coloured purple) the bands that
would be found if the bound-state formation rate were a factor 2 larger than our calculations,
as might arise from higher-order QCD or other effects~\footnote{We evaluate the $\alpha_s$ appearing in the 
Sommerfeld enhancement factor at a scale $\beta m_{\tilde g}$ that is typical of the momentum transfer of the 
soft-gluon exchanges responsible for the Sommerfeld effect~\cite{bcg98}, and take $\beta = 0.3$, 
which is comparable to the thermal velocities of the gluinos at the freeze-out temperature. 
Because the Sommerfeld enhancement is a precursor to the formation of bound states, 
for simplicity we take the same $\alpha_s$ in evaluating the bound-state effects. 
The full QCD potential and the thermal mass of the gluon were considered in computing the 
Sommerfeld enhancement in~\cite{deSimone:2014pda}, and they should be relevant
also to the calculation of the bound-state effects. We estimate that these effects could result in a shift of our 
orange bands inward by $\sim$ 10\%, based on a comparison of the light green and dark green bands in the 
lower right panel of Fig. 2 in~\cite{deSimone:2014pda}, and we expect a similar shift for our black bands. 
However, these two additional effects could be compensated by contributions from excited bound states
that we do not include. We note that it was estimated in~\cite{excitedbs} in the context of a hidden-sector stau that 
excited bound states can at most introduce a factor of $\sim$ 1.6 in the thermally-averaged bound-state 
formation cross section, compared to that obtained by considering only the ground state, 
but the color charge in the gluino case may change this value. We therefore plot purple bands 
with an uncertainty of a factor 2 to allow for these uncertainties.}.

The upper left panel of Fig.~\ref{fig:bino} is for the case $m_{\tilde q}/m_{\tilde g} = 1.1$,
where the $t$ and $u$ channels partially cancel the $s$-channel contributions to the
gluino annihilation cross section. Here
we see that the black band calculated including both
the Sommerfeld enhancement and gluino bound-state effects extends to $m_\chi \sim 6.2$ to 6.4~TeV.
In this case, the numerical effects of the Sommerfeld enhancement are similar to those of gluino bound-state formation, and
both effects are considerably larger than the current observational uncertainties in the dark matter density represented by
the breadths of the bands. The purple band, which includes an allowance of a factor 2
uncertainty in the bound-state effects, as might arise from higher-order
QCD, excited bound states, etc., extends to larger $m_\chi \sim 7.2$ to 7.5~TeV.
In the case $m_{\tilde q}/m_{\tilde g} = 10$ (upper right panel of Fig.~\ref{fig:bino}),
the effect of bound-state formation is somewhat smaller than the Sommerfeld effect, and the black (purple) band
extends to $m_\chi \sim 8$ (9)~TeV. These trends are also seen in the case
$m_{\tilde q}/m_{\tilde g} = 50$ (lower left panel of Fig.~\ref{fig:bino}), where the black and purple bands
also extend to $m_\chi \sim 8$ (9)~TeV. On the other hand, the results for $m_{\tilde q}/m_{\tilde g} = 120$
(lower right panel of Fig.~\ref{fig:bino}) are quite different. The Sommerfeld effect is much larger
than the bound-state effect though the latter is still slightly larger than the widths of the coloured bands corresponding to the 3-$\sigma$ ranges for the cold dark matter density. 
Also, the allowed range of 
the LSP mass is greatly reduced, extending only to $\sim 6.1 \tev$ ($\sim 6.3 \tev$ allowing for a factor 2
uncertainty in the bound-state effects).

\begin{figure}
\begin{center}
\begin{tabular}{c c}
\hspace{-0.6cm}
\includegraphics[height=5.5cm]{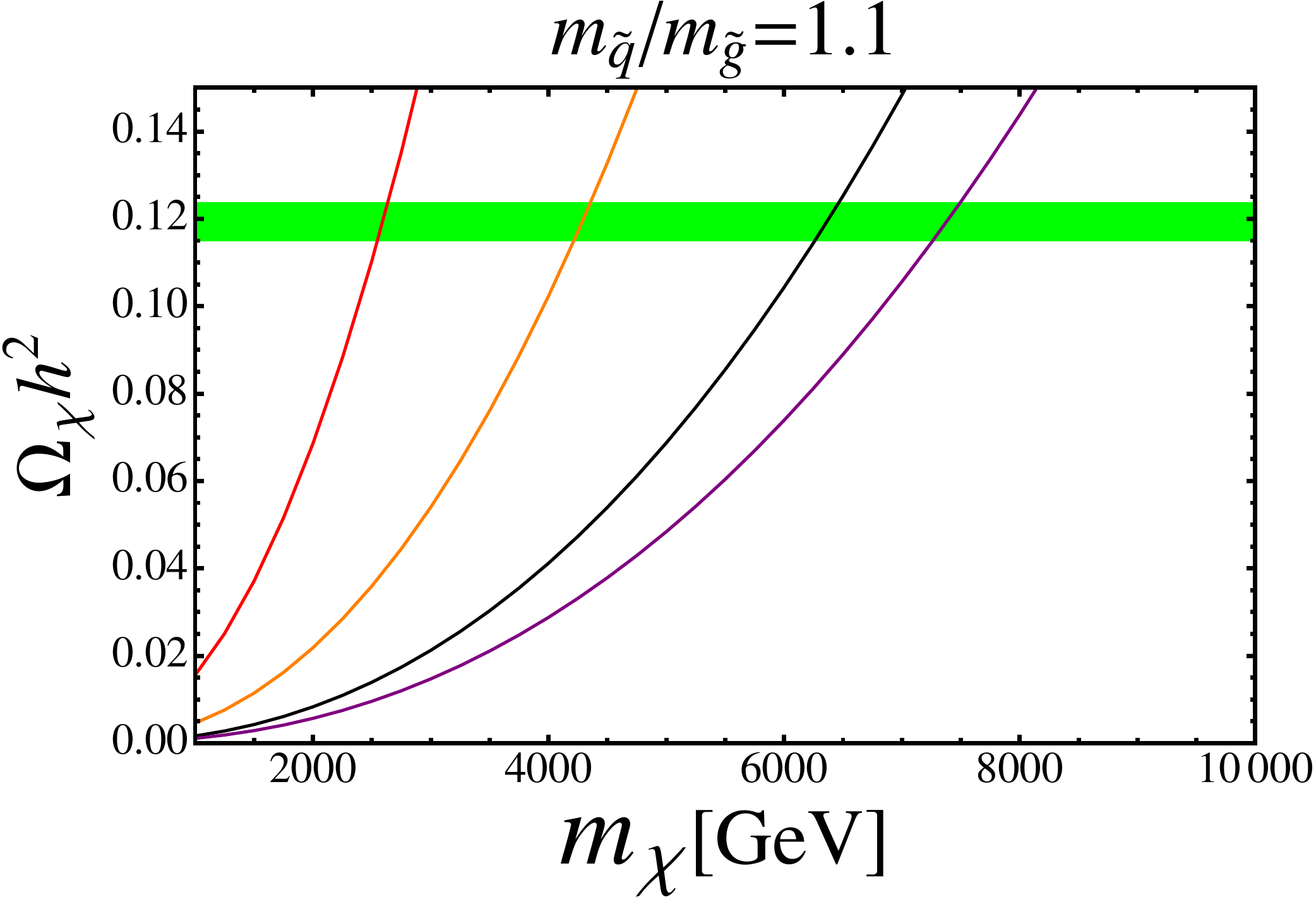} & 
\hspace{-0.6cm}
\includegraphics[height=5.5cm]{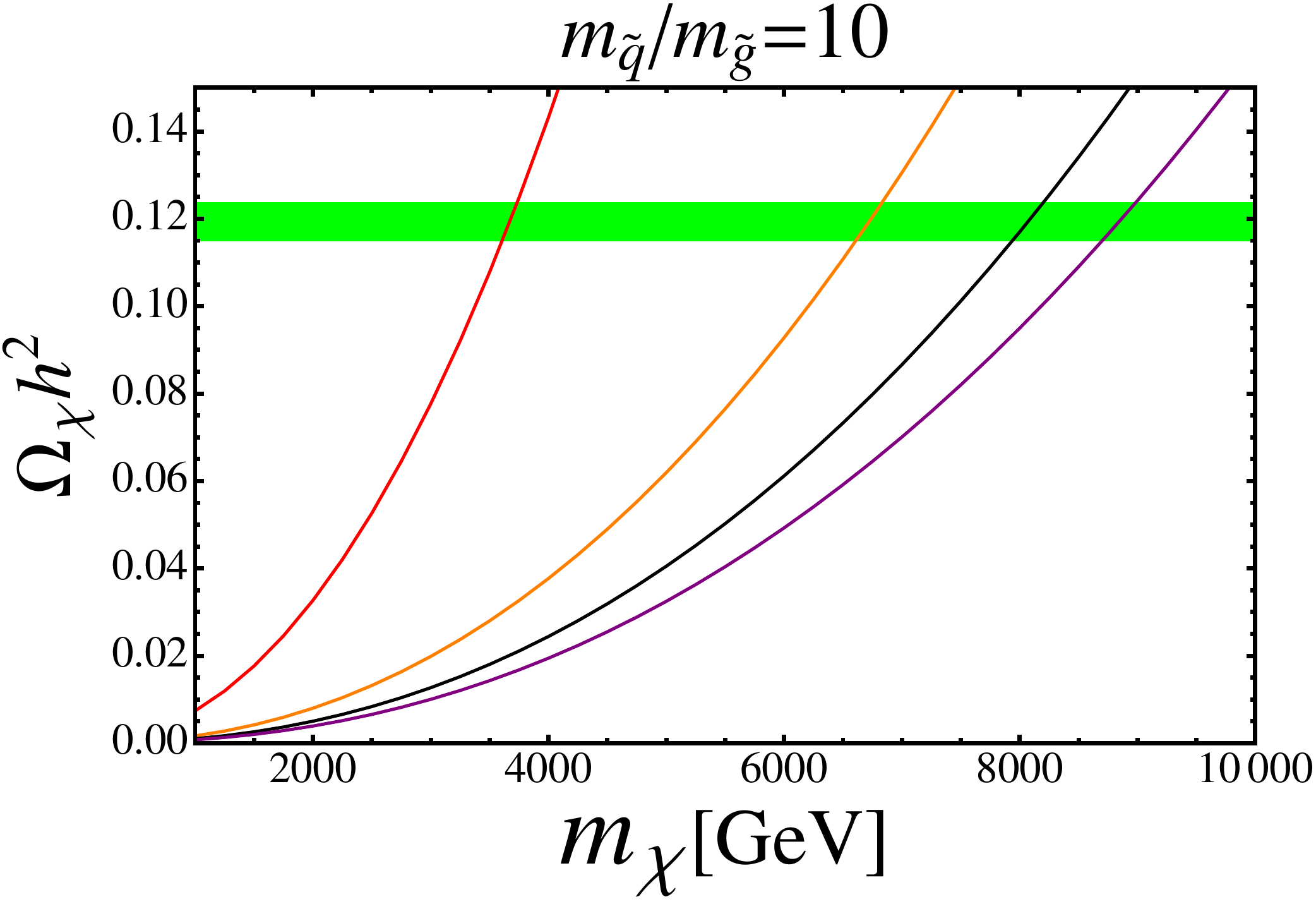} \\
\hspace{-0.6cm}
\includegraphics[height=5.5cm]{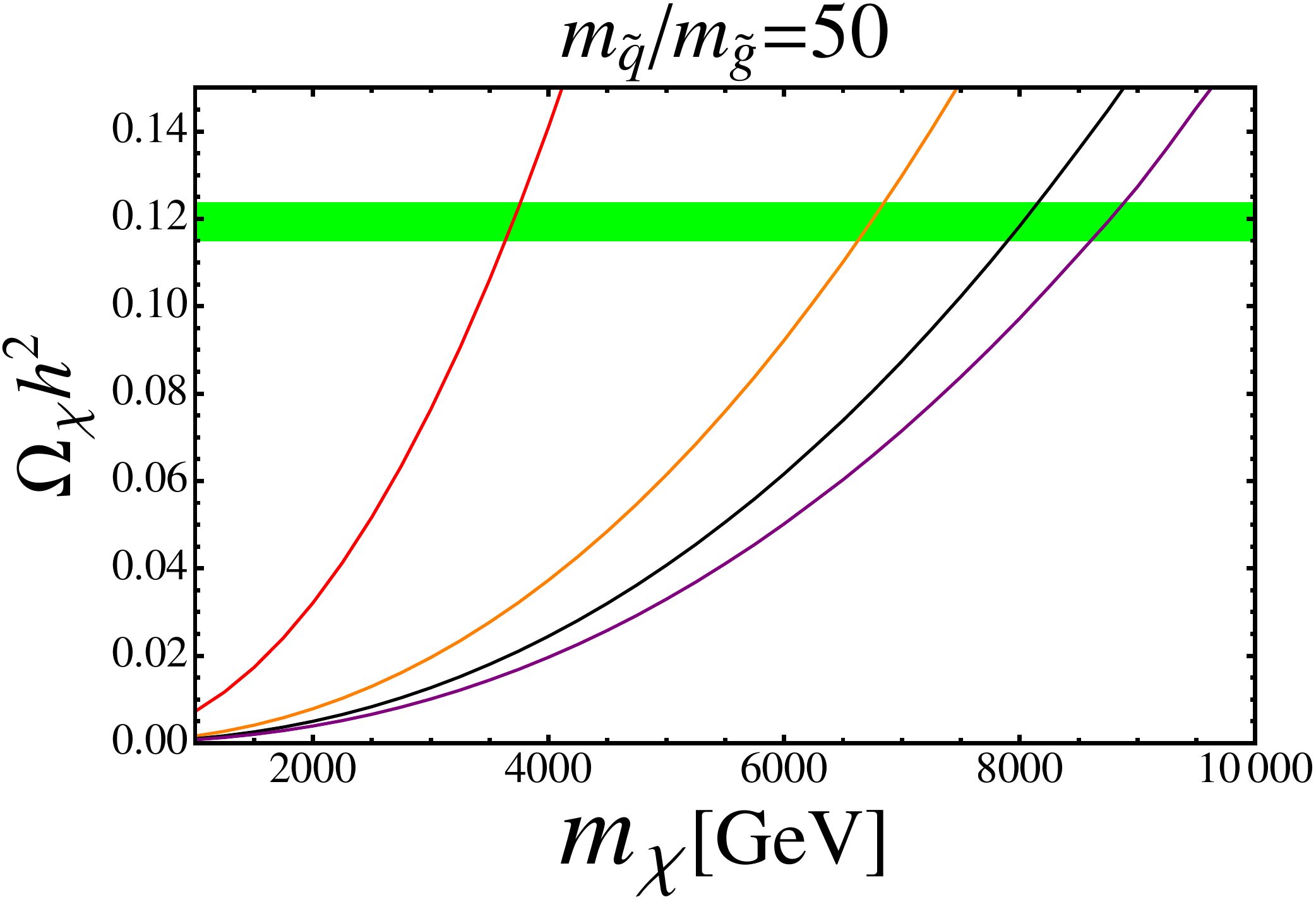} &
\hspace{-0.6cm}
\includegraphics[height=5.5cm]{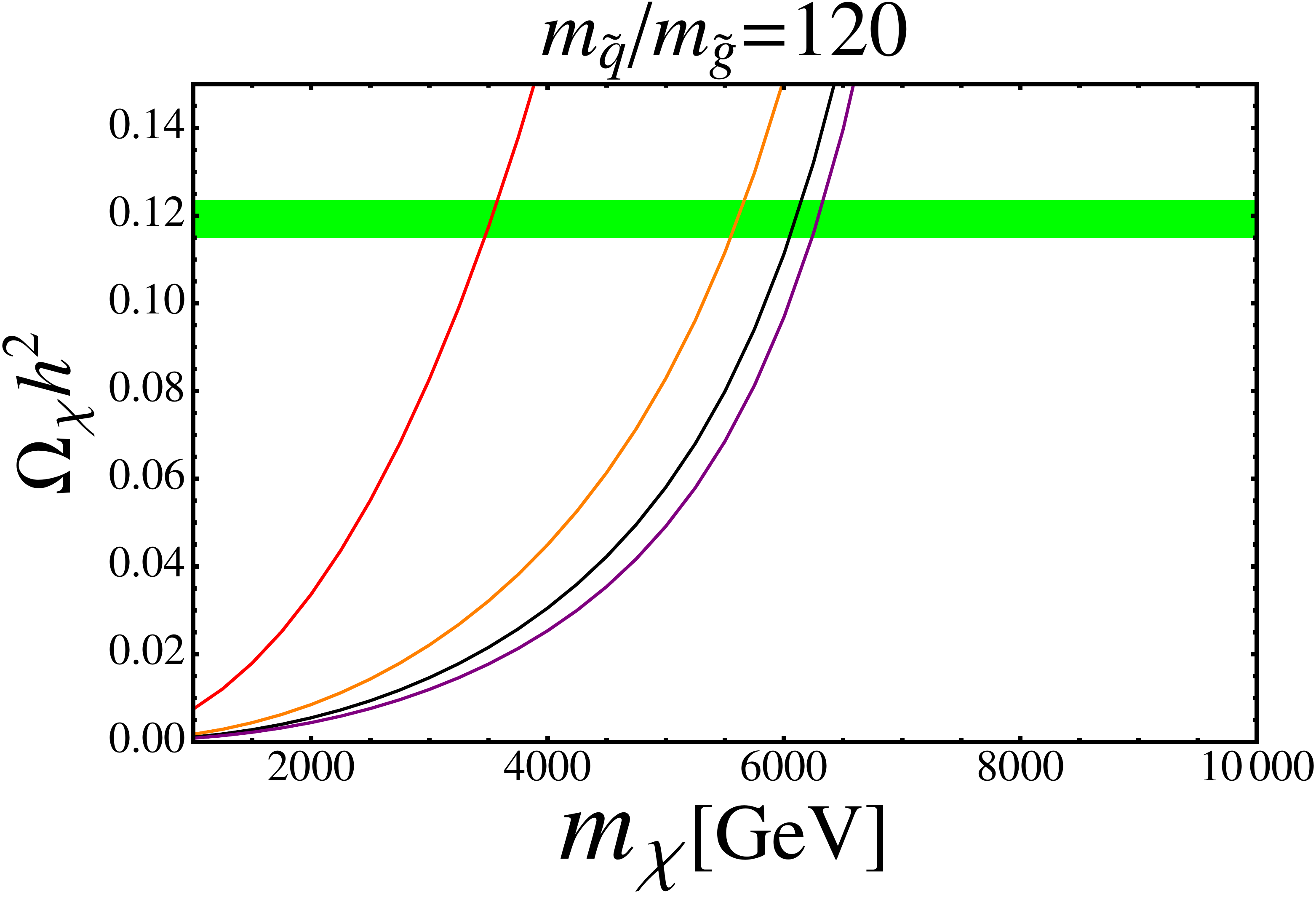}
\end{tabular}
\end{center}   
\caption{\label{fig:bino2}\it
The locations of the endpoints of the gluino coannihilation strips for different values of $\ohsq$,
using the same colour conventions as in Fig.~\protect\ref{fig:bino}. As in that Figure, the calculations
assume the different values $m_{\tilde q}/m_{\tilde g} = 1.1$
(upper left), $10$ (upper right), $50$ (lower left) and $120$ (lower right).
The horizontal green bands show the 3-$\sigma$ band $0.1151 < \ohsq < 0.1235$.
}
\end{figure}

Fig.~\ref{fig:bino2} displays these effects differently, exhibiting the positions
of the endpoints ($m_{\tilde g} = m_\chi$) of the gluino coannihilation strips as functions of the assumed value of $\ohsq$,
again from calculations with neither the Sommerfeld enhancement nor gluino bound states (red lines),
with the Sommerfeld enhancement but without the bound states (orange lines), with both effects
included (black lines), and allowing for a factor 2
uncertainty in the bound-state effects (purple lines). The horizontal green bands again show the
3-$\sigma$ band $0.1151 < \ohsq < 0.1235$. We recall that lower values of $\ohsq$ would be relevant if the LSP provides
only part of the dark matter density, e.g., if there is also a significant axion component, and parameter choices yielding higher
values of $\ohsq$ in conventional Big Bang cosmology (as assumed here)
could be relevant in models with non-standard cosmological
evolution~\footnote{It is worth recalling that for $m_{\tilde g} \sim 5$~TeV
the bound-state binding energy is ${\cal O}(50)$~GeV and the freeze-out temperature in conventional
Big Bang cosmology is hundreds of GeV, so the assumption made in this paper of
standard cosmological evolution is rather different from that made in more conventional thermal dark
matter scenarios where the freeze-out temperature may be in the GeV range.}.
As was also seen in Fig.~\ref{fig:bino}, the smallest value of $m_{\tilde q}/m_{\tilde g} = 1.1$
(upper left) leads to smaller values of $m_\chi$ for any fixed value of $\ohsq$,
as compared to the $m_{\tilde q}/m_{\tilde g} = 10$ case (upper right).
The choice $m_{\tilde q}/m_{\tilde g} = 50$ (lower left) leads to a marginally smaller value of $m_\chi$,
and the choice $m_{\tilde q}/m_{\tilde g} = 120$ (lower right) leads to significantly lower
values of $m_\chi$ for any fixed value of $\ohsq$. The effect of including bound-state effects is
to increase the range of $m_\chi$ compatible with the measured value of $\ohsq$ by $\sim 50$\%
for $m_{\tilde q}/m_{\tilde g} = 1.1$, decreasing to $\sim 20$\% for $m_{\tilde q}/m_{\tilde g} = 10$ to $50$.

\begin{figure}[ht!]
\begin{center}
\includegraphics[height=9cm]{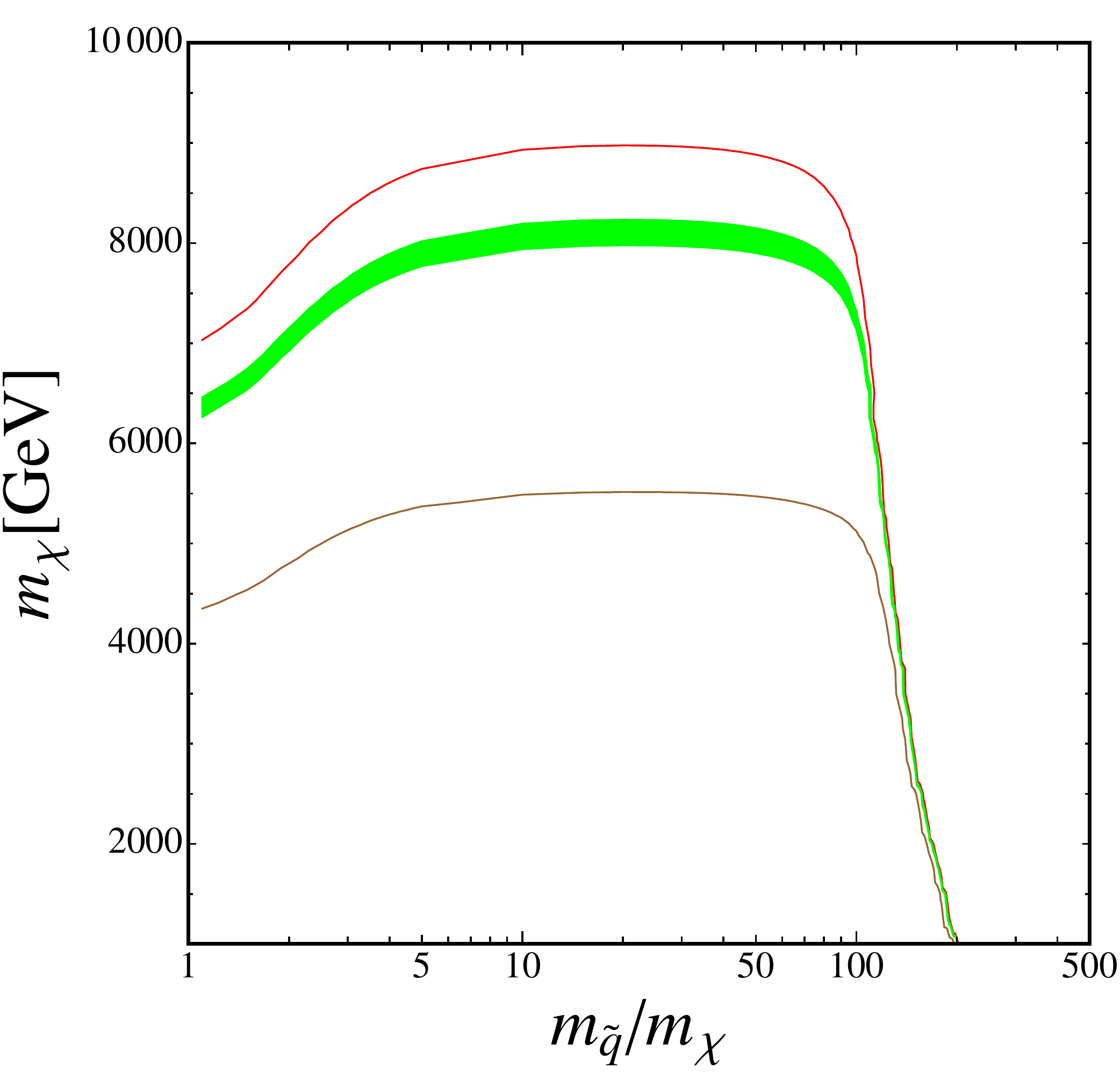}
\end{center}   
\caption{\label{fig:nisab}\it
The value of $m_\chi$ at the endpoint 
of the gluino coannihilation strip when $\Delta m = m_{\tilde g} - m_\chi = 0$ in the Bino LSP case,
as a function of $m_{\tilde q}/m_{\tilde g}$. The drop at small $m_{\tilde q}/m_{\tilde g}$
is due to the cancellations between the $s$-, $t$-
and $u$-channel diagrams for gluino pair annihilation, and that at large $m_{\tilde q}/m_{\tilde g}$ is due to the
decoupling of the gluino and neutralino densities. The green band corresponds to the current $3$-$\sigma$ range of
the dark matter density: $\Omega_\chi h^2 = 0.1193 \pm 0.0042$, and the brown and red contours are for $\Omega_\chi h^2 = 0.05$ and $0.15$,
respectively.
}
\end{figure}

Finally, we show in Fig.~\ref{fig:nisab} the value of $m_\chi$ at the endpoint 
of the coannihilation strip when $\Delta m = 0$ and $\Omega_\chi h^2 = 0.1193 \pm 0.0042$ (green band), as a function of $m_{\tilde q}/m_{\tilde g}$:
the brown and red contours are for $\Omega_\chi h^2 = 0.05$ and $0.15$, respectively.
The band and contours exhibit the inverse of the behaviour of the relic density  seen previously in Fig.~\ref{fig:basin}.
The neutralino mass at low $m_{\tilde q}/m_{\tilde g}$ is below the maximum value of $m_\chi$,
which has a shallow maximum around $m_{\tilde q}/m_{\tilde g} = 10$ to $50$, and falls sharply when $m_{\tilde q}/m_{\tilde g}
\gtrsim 100$, reflecting the effect of a breakdown in ${\tilde g} - \chi$ conversion.
We conclude that, within the framework studied
here, $m_\chi \lesssim 8$~TeV (rising to $\sim 9$~TeV when allowing for a factor 2 uncertainty in the bound-state formation rate)
in the Bino LSP case.

\subsection{Wino LSP}

\begin{figure}
\begin{center}
\begin{tabular}{c c}
\hspace{-0.6cm}
\includegraphics[height=7cm]{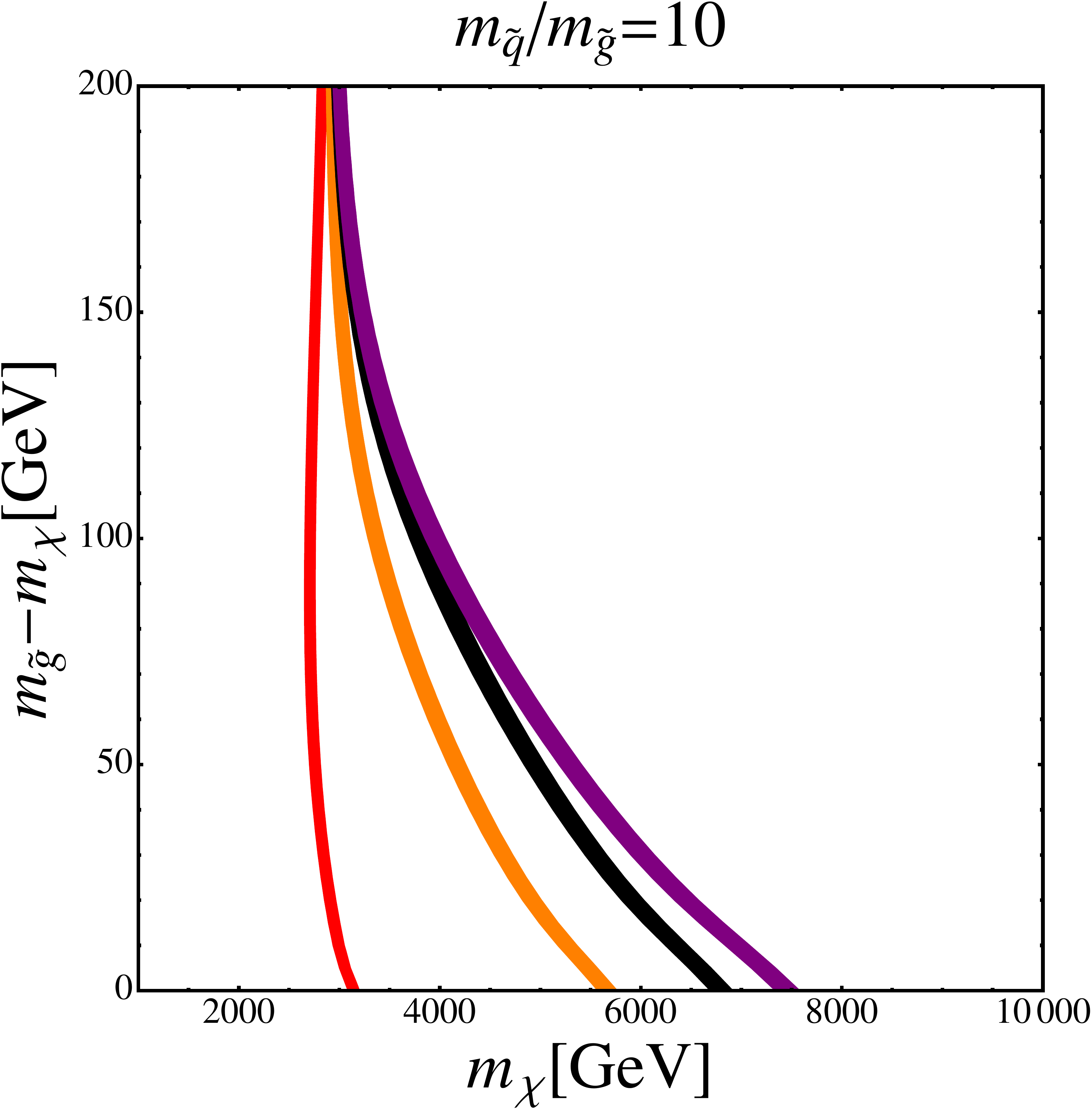} & 
\hspace{-0.6cm}
\includegraphics[height=7cm]{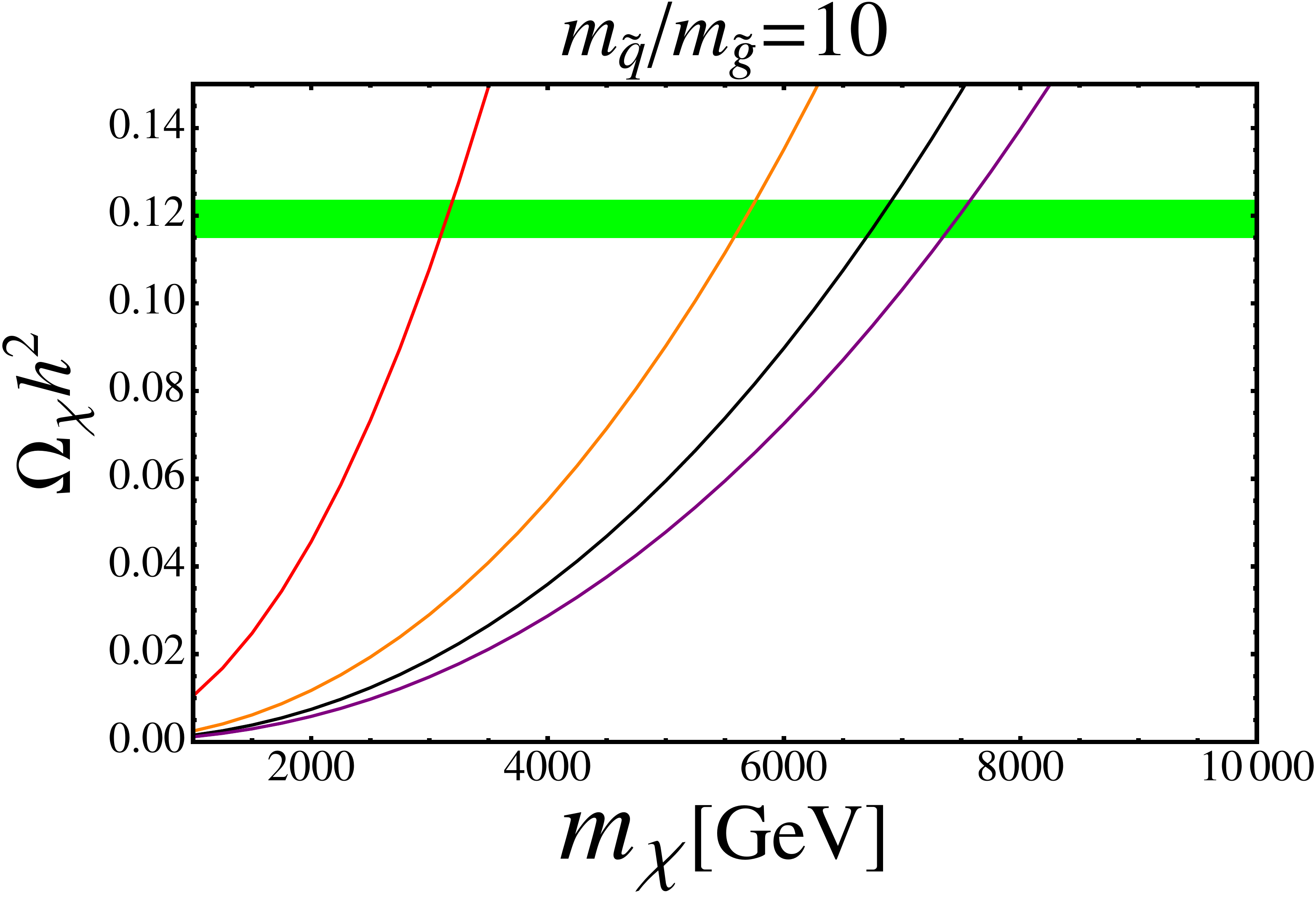} \\
\end{tabular}
\end{center}   
\caption{\label{fig:wino}\it
As in Fig.~\protect\ref{fig:bino} and~\protect\ref{fig:bino2}, but for a Wino LSP.
}
\end{figure}

We now consider the case of a Wino LSP. The left panel of Fig.~\ref{fig:wino} displays
the gluino-Wino coannihilation strips for $\ohsq = 0.1193 \pm 0.0042$
for $m_{\tilde q}/m_{\tilde g} = 10$, using the same colour
codings as for the Bino case (red with neither the Sommerfeld enhancement nor
gluino bound states, orange including the QCD Sommerfeld enhancement but again no
bound-state effects, black with both effects included, and purple with the
bound-state formation rate enhanced by a factor 2). We see that in this case the black
coannihilation strip extends to $m_\chi \sim 7$~TeV. Note that the curves appear to diverge
at low $m_\chi$. The reason is that even in the absence of coannihilation (large $m_{\tilde g} - m_\chi$),
Wino-Wino annihilations are strong enough to suppress the relic density below the density indicated
by Planck and other experiments when
$m_\chi \la 3$ TeV~\footnote{The Sommerfeld enhancements in the calculations of the thermal relic abundance of a 
Wino- or Higgsino-like LSP were discussed in detail in~\cite{winohiggsinorelicwsom}. 
We have included an estimate of the Sommerfeld enhancement factor for the Wino-Wino annihilations, 
and we get a similar result as~\cite{gluino2014}, namely $m_\chi \sim 3$~TeV,
in the limit where the effect of gluino coannihilation is absent. We do not include the 
Sommerfeld enhancement for the Higgsino-Higgsino annihilations, 
since its effect is much milder in the relic abundance calculations, as shown in the literature.}. 
The right panel of 
Fig.~\ref{fig:wino} shows how $\ohsq$ at the endpoints of the strips
varies with $m_\chi$. As previously, the colours of the lines correspond to
the colours of the strips in the left panel of Fig.~\ref{fig:wino}. We see that 
the black line crosses the horizontal green band where $\ohsq = 0.1193 \pm 0.0042$
for $m_\chi \sim 7$~TeV.

\begin{figure}
\begin{center}
\includegraphics[height=7.5cm]{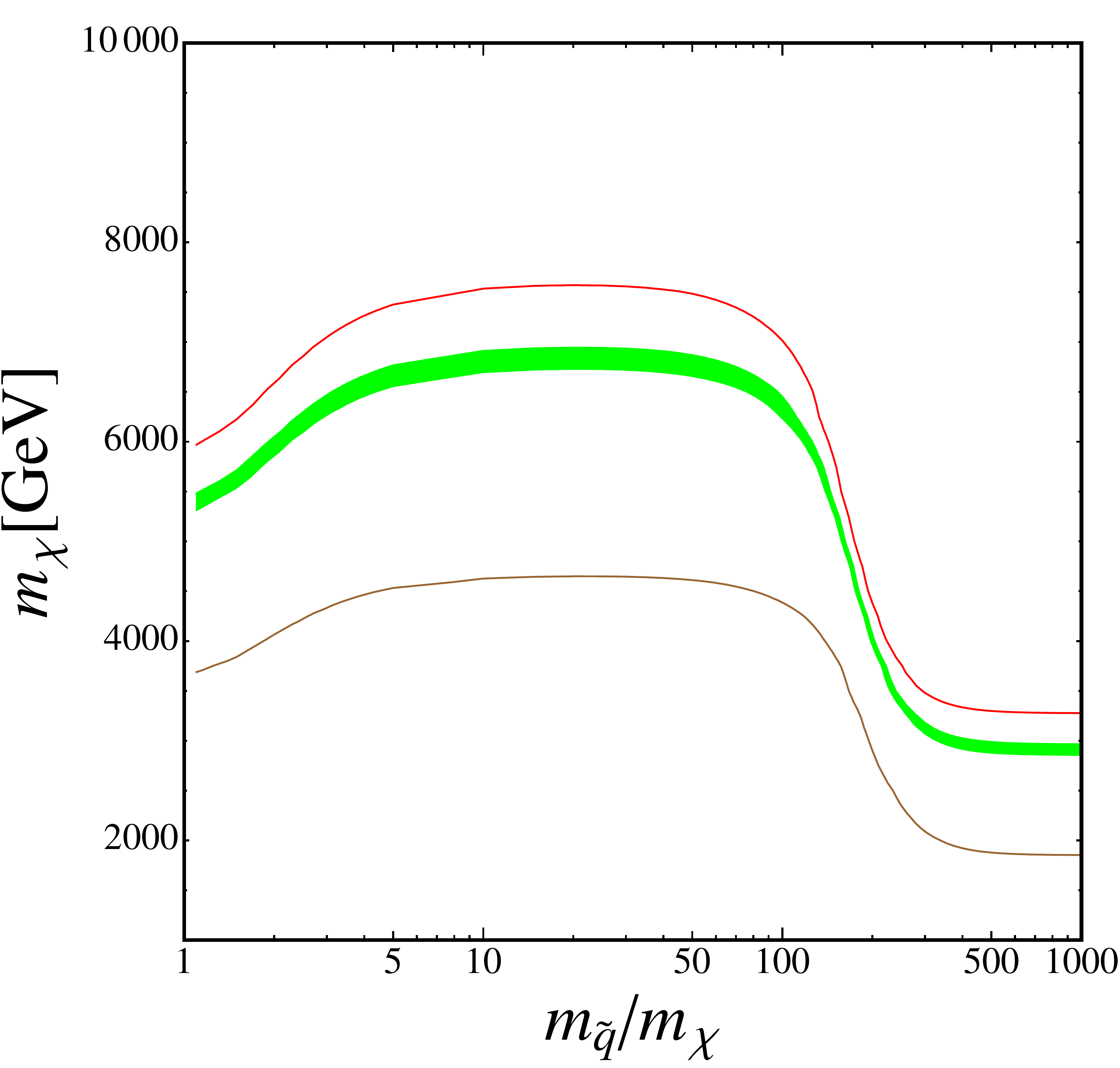}
\includegraphics[height=7.5cm]{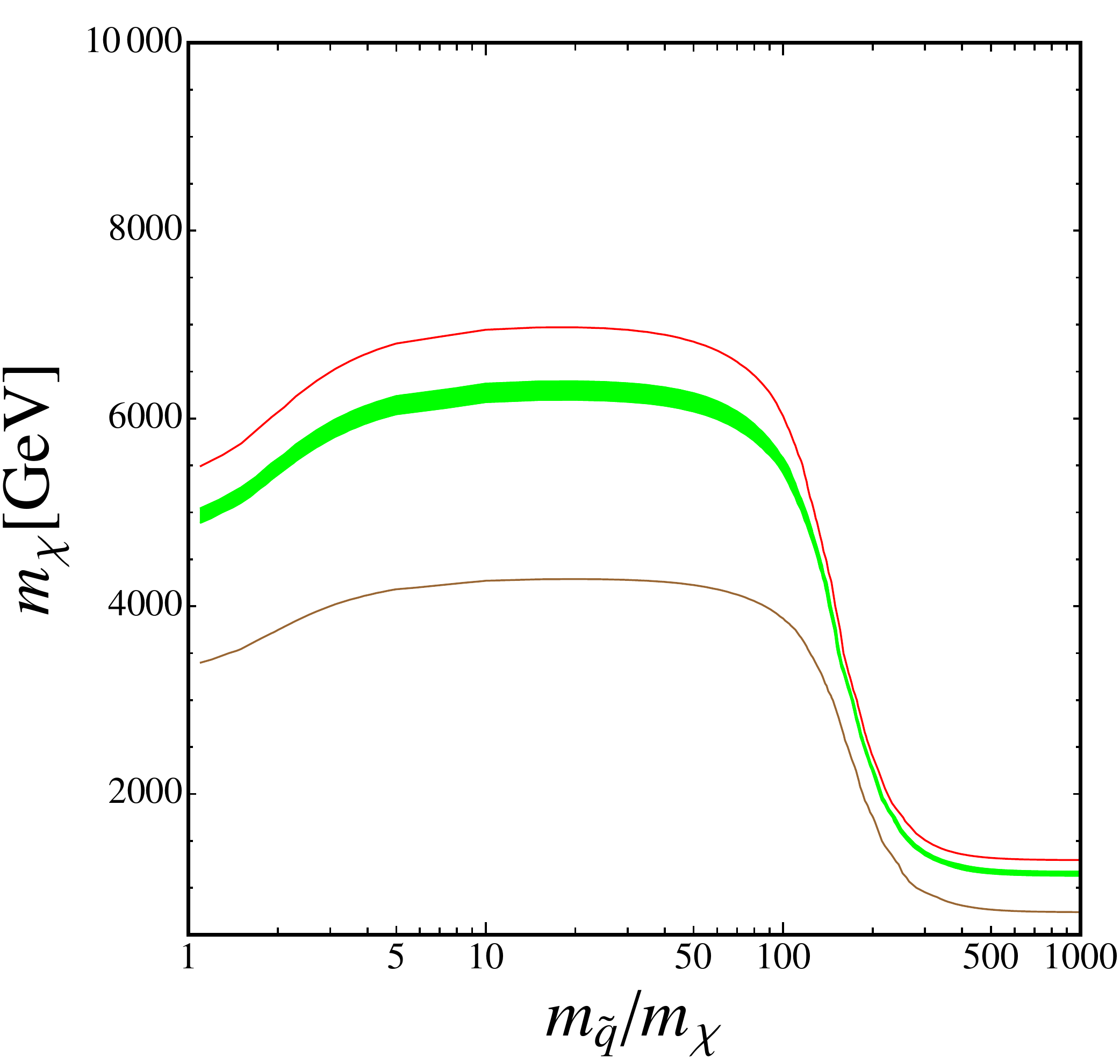}
\end{center}   
\caption{\label{fig:nisabWH}\it
As in Fig.~\protect\ref{fig:nisab}, but for a Wino LSP (left panel) and a Higgsino LSP (right panel). 
}
\end{figure}

The left panel of Fig.~\ref{fig:nisabWH} is the analogue of Fig.~\ref{fig:nisab} for the case of
a Wino LSP, with the green band corresponding to $\ohsq = 0.1193 \pm 0.0042$
and the brown and red contours to $\Omega_\chi h^2 = 0.05$ and $0.15$.
We see that $\ohsq$ is within the preferred range for $m_\chi \sim 7$~TeV
over a broad range $5 \lesssim m_{\tilde q}/m_{\tilde g} \lesssim 100$. The percentage
increase in the allowed range of $m_\chi$ due to bound-state effects, as a function of $m_{\tilde q}/m_{\tilde g}$, is similar
to the Bino case. As shown in Fig.~\ref{fig:nisab} for the Bino case, the fall in the $\ohsq$ to lower values of $m_\chi$
is due to the breakdown of ${\tilde g} - \chi$ conversion. The curve hits a  plateau for
$m_{\tilde q}/m_{\tilde g} \gtrsim 300$ which represents the decoupling limit at $m_\chi \sim 3$ TeV.

\subsection{Higgsino LSP}

\begin{figure}
\begin{center}
\begin{tabular}{c c}
\hspace{-0.6cm}
\includegraphics[height=7cm]{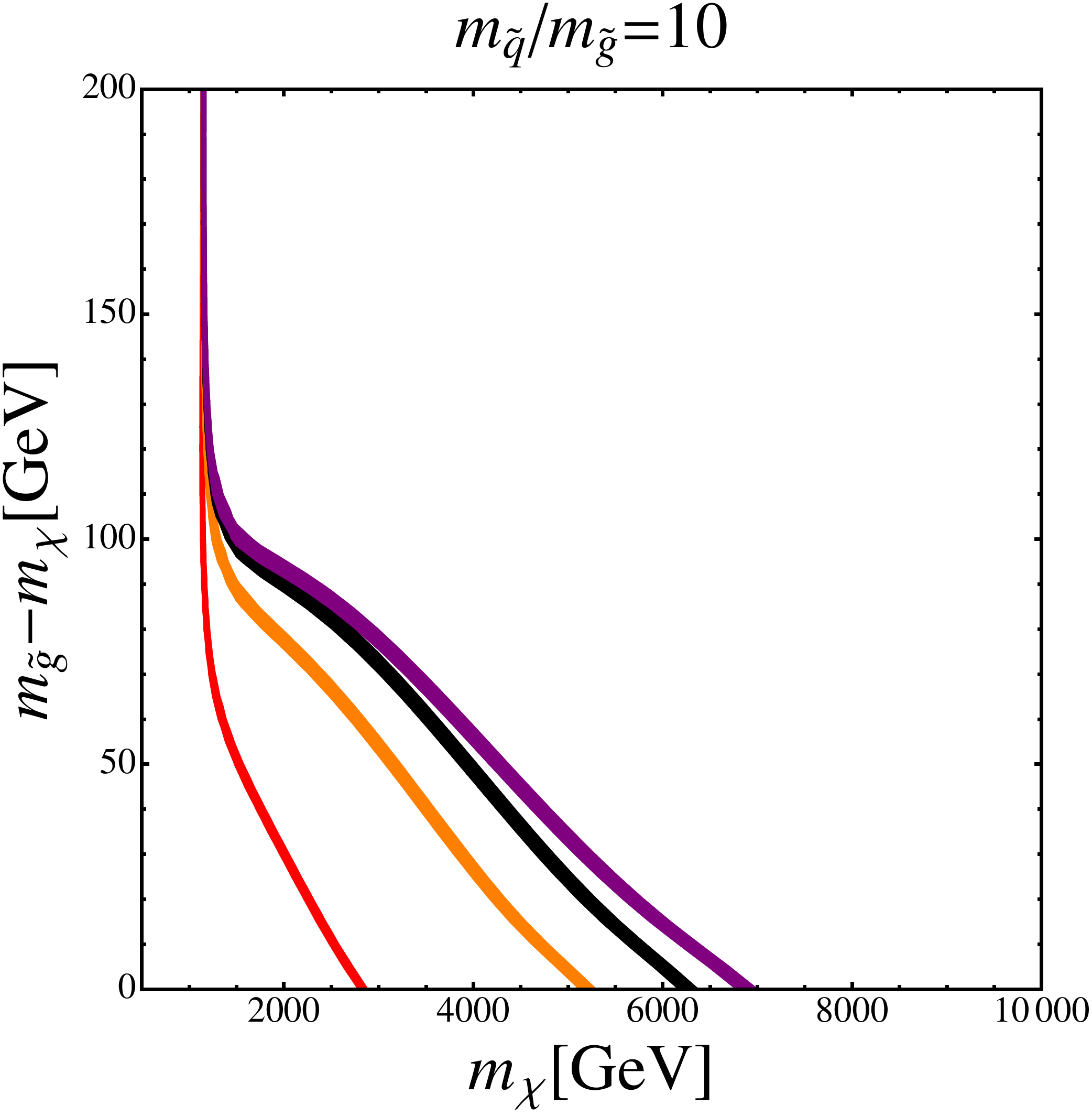} & 
\hspace{-0.6cm}
\includegraphics[height=7cm]{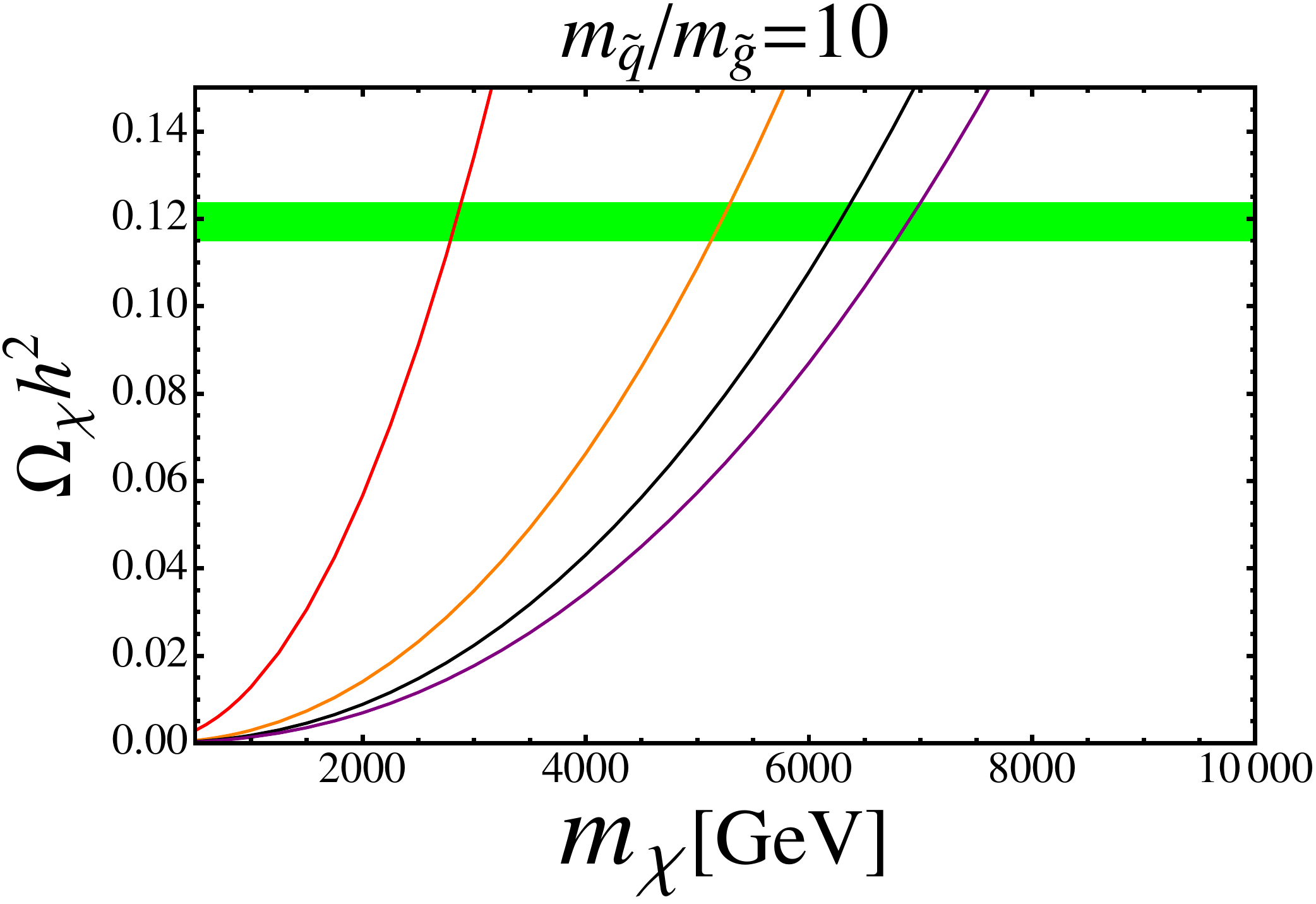} \\
\end{tabular}
\end{center}   
\caption{\label{fig:Higgsino}\it
As in Fig.~\protect\ref{fig:bino} and~\protect\ref{fig:bino2}, but for a Higgsino LSP.
}
\end{figure}

We now consider the case of a Higgsino LSP. The left panel of Fig.~\ref{fig:Higgsino} displays
the gluino-Higgsino coannihilation strips for $\ohsq = 0.1193 \pm 0.0042$ and $m_{\tilde q}/m_{\tilde g} = 10$ using the same colour
codings as for the Bino and Wino cases (red with neither the Sommerfeld enhancement nor
gluino bound states, orange including the QCD Sommerfeld enhancement but again no
bound-state effects, black with both effects included, and purple with the
bound-state formation rate enhanced by a factor 2). In this case the black
strip extends to $m_\chi \sim 6$~TeV at the endpoint where $\Delta m = 0$. 
The Higgsino couplings depend on $\tan \beta$ and in our calculations we have taken $\tan \beta = 10$. 
Our results are very weakly dependent on this choice.
Once again we see a divergence of the contours at low $m_\chi$.  In this case, when $m_\chi \la 1.2$ TeV,
Higgsino-Hiiggsino annihilations are sufficient to reduce the relic density below the Planck density. The right panel of 
Fig.~\ref{fig:Higgsino} shows how $\ohsq$ at the endpoints
varies with $m_\chi$, with the colours of the lines corresponding again to
the colours of the strips in the left panel of Fig.~\ref{fig:Higgsino}. 
The black line crosses the horizontal green band where $\ohsq = 0.1193 \pm 0.0042$
for $m_\chi \sim 6$~TeV. As seen in the right panel of Fig.~\ref{fig:nisabWH}, similar
values of $m_\chi$ are found for a range $5 \lesssim m_{\tilde q}/m_{\tilde g} \lesssim 100$,
with the drops in the $\ohsq$ contours to lower values of $m_\chi$
again being due to cross section cancellations at low $m_{\tilde q}/m_{\tilde g}$ and due to the breakdown of ${\tilde g} - \chi$ conversion at high $m_{\tilde q}/m_{\tilde g}$. As in the case of the Wino,
the curves drop to a plateau for
$m_{\tilde q}/m_{\tilde g} \gtrsim 300$ representing the decoupling limit. 
In this case, the asymptotic value of $m_\chi$ is $\sim 1.2$ TeV.  The percentage
increase in the allowed range of $m_\chi$ due to bound-state effects is again similar
to the Bino case. 

The decreases in the maximum values of $m_\chi$ allowed in the Wino and Higgsino cases,
compared to the Bino case, are due to the effect noted in~\cite{hep-ph/0609290}, namely
that coannihilations may, under some circumstances, {\it increase} the relic abundance by
coupling `parasitic' degrees of freedom. In the Bino (Wino) (Higgsino) case, there are 2 (6) (8)
electroweak degrees of freedom, linked by coannihilation to the gluinos, that contribute incrementally to
the relic abundance. This effect is compensated by the decreases in the maximum values of $m_\chi$
that we find in the Wino and Higgsino cases.

\section{Summary}

We have studied in this paper MSSM scenarios in which the LSP is (almost)
degenerate with the gluino, exploring the characteristics and locating the endpoints
of the gluino-LSP coannihilation strip in the cases where the LSP is the Bino,
the neutral Wino or a neutral Higgsino. Important ingredients in our analysis
are the Sommerfeld enhancement of gluino annihilation rates, gluino-gluino
bound-state formation and gluino-neutralino conversion. As we show, these can affect significantly the preferred
range of the gluino-LSP mass difference along the coannihilation strip, and also
the position of the endpoint.

In the Bino LSP case, we find that at the endpoint the LSP mass $\sim 8$~TeV,
increasing to $\sim 9$~TeV if we allow for a factor 2 increase in the bound-state
formation rate above our calculations. These values are decreased by $\sim 1$~TeV
if the LSP is a Wino, and by a further $\sim 1$~TeV if it is a neutral Higgsino.
The upper limit on the LSP mass of $\sim 8$~TeV is weakly sensitive to the squark
mass for $10 \la m_{\tilde q}/m_{\tilde g} \la 50$, but is substantially reduced for either
smaller or larger values of $m_{\tilde q}/m_{\tilde g}$. In all cases, the percentage
increase in the allowed range of $m_\chi$ due to bound-state effects may be as large as 50\%.

We are loath to claim that our upper limit on the LSP is absolute, but we do note
that it is substantially higher than what is possible along the stop coannihilation strip,
reflecting the larger annihilation rates that are possible for the gluino because of its
larger colour charge. However, these annihilation rates also depend on the masses
of other sparticles, notably the squarks in the gluino NLSP case studied here. As have
shown, the decrease in upper limit on the LSP mass for small $m_{\tilde q}/m_{\tilde g}$
is due to cancellations in the annihilation matrix elements, whilst the decrease at large
$m_{\tilde q}/m_{\tilde g}$ is due to the breakdown of gluino-LSP conversion. However,
we have not studied the limit $m_{\tilde q}/m_{\tilde g} \to 1$, where many more coannihilation
processes would come into play, as might also be the case in non-minimal supersymmetric models.

Nevertheless, our analysis does show that a large mass reach to at least 8~TeV
will be necessary to explore conclusively the possibility of supersymmetric dark matter
within the MSSM and a conventional cosmological framework.

\section*{Acknowledgements}

The authors would like to thank Brian Batell, Mathias Garny, Matthew McCullough and Hua-Sheng Shao 
for very helpful discussions.
The work of J.E. was supported in part by the London Centre for Terauniverse Studies
(LCTS), using funding from the European Research Council via the Advanced Investigator
Grant 267352 and from the UK STFC via the research grant ST/J002798/1. The work of F.L.
was also supported by the European Research Council Advanced Investigator
Grant 267352. The work of K.A.O.
was supported in part by DOE grant DE-SC0011842 at the University of Minnesota.

\section*{Appendix A}
In this appendix, we first recall our procedure for computing the 
thermal-averaged velocity-weighted cross-sections $\langle\sigma_{12} v_{\rm rel}\rangle$
for the process $1+2\to 3+4$ as is necessary for solving the Boltzmann equations in
Section 5 in an efficient manner, extending the approach used in {\tt SSARD}~\cite{SSARD}.
More details of our approach can be found in~\cite{eos,efosi}.
 
We start with the squared transition matrix elements $\tsq$
(summed over final spins and colors, averaged over initial spins and colors) for the coannihilation processes of interest,
which are given here in 
Appendix B, expressed
as functions of the Mandelstam variables $s$, $t$, $u$.
We then express $\tsq$ in terms of $s$ and the
scattering angle $\tcm$ in the centre-of-mass frame, as described 
in~\cite{efosi}. Next we define
\begin{eqnarray}
w(s) &\equiv& {1\over4}
              \int {d^3 p_3\over(2\pi)^3 2 E_3}\,{d^3 p_4\over(2\pi)^3 2 E_4}
              \,(2\pi)^4\delta^4(p_1+p_2-p_3-p_4)\; \tsq
\nonumber \\
&=& {1\over32\pi}\,{p_3(s)\over s^{1/2}} \int_{-1}^{+1}d\cos\tcm\,\tsq \;.
\label{w}
\end{eqnarray}  
The total annihilation cross section
$\sigma_{12}(s)$ is given in terms of $w(s)$ by
$\sigma_{12}(s) = w(s)/s^{1/2}p_1(s)$~\footnote{Our $w(s)$
is also the same as $w(s)$ in~\cite{swo,fkosi,efo},
which is written as $W/4$ in~\cite{eg}.}.

The usual partial-wave 
expansion can be obtained by 
expanding $\tsq$ in powers of $p_1(s)/m_1$.
The odd powers vanish upon
integration over $\tcm$, while the zeroth- and second-order terms
correspond to the usual $s$ and $p$ waves, respectively.
We can therefore evaluate the $s$- and $p$-wave contributions to
$w(s)$ simply by evaluating $\tsq$
at two different values of $\cos\tcm$.

The proper procedure for thermal averaging has been discussed
in~\cite{swo,gg} for the case $m_1=m_2$, and
in~\cite{fkosi,eg} for the case $m_1\ne m_2$, so we do not discuss
it in detail here. One finds the coefficients $a$ and $b$ in the expansion (\ref{ab})
of the thermal-averaged cross-sections for the processes of interest:
\begin{equation}
\langle\sigma_{12} v_{\rm rel}\rangle
= a_{12} + b_{12} \, x^{-1} + {\cal O}(x^{-2}) \;,
\label{sv3}
\end{equation}
where $x \equiv m_1/T$ (assuming $m_1<m_2$)
by following the prescription given in \cite{efosi}, using the
transition amplitudes
listed in Appendix B for each final state.   
When the conversion rates are large compared to the Hubble rate, these amplitudes can be used to compute the total effective coefficients
$a_{\rm eff}$ and $b_{\rm eff}$ by performing the sum over initial
states as in (\ref{sv2}), and we then integrate the rate equation (\ref{rate2}) 
numerically to obtain the relic density.

\section*{Appendix B}

The $2 \to 2$ (co)annihilation processes relevant to the gluino-neutralino (and, in the Wino and Higgsino
cases, gluino-charginos) system are 
$\tilde{g} \tilde{g} \to g g$, 
$\tilde{g} \tilde{g} \to q_A \bar{q}_B$,
$\tilde{g} \chi^0_i \to q_A \bar{q}_B$
and 
$\tilde{g} \chi^+_j \to u_A \bar{d}_B$,
where the indices $A, B = 1, 2, 3$ for three generations, the neutralino index $i = 1, ..., 4$ and the chargino index $j = 1, 2$. 

Here we list the $\tsq$ for each of these processes, separating the contributions
from $s$-, $t$- and $u$-channel diagrams. In the following expressions, final spins and colors are summed over, 
and initial spins are averaged over. A factor $c_{ini}$ is used to average over initial colors. Therefore, $\tsq$ takes the form 
\beq
\tsq = c_{ini} ({\cal T}_s \! \times \! {\cal T}_s + {\cal T}_t \! \times \! {\cal T}_t + {\cal T}_u \! \times \! {\cal T}_u 
+ {\cal T}_s \! \times \! {\cal T}_t + {\cal T}_s \! \times \! {\cal T}_u + {\cal T}_t \! \times \! {\cal T}_u) \,.
\eeq 
We note that there is also the charge-conjugated process for the chargino, $\tilde{g} \chi^-_j \to \bar{u}_A d_B$,
which we do not list separately.  

\subsection*{$\tilde{g} \tilde{g} \to g g$}

There is an $s$-channel gluon-exchange diagram, and $t$- and $u$-channel gluino-exchange diagrams. 
We note that, because there are two identical gluons in the final state, an extra factor of $1/2$ is needed when performing the momentum integration in (\ref{w}).

We find
\begin{eqnarray}
c_{ini} &=& \frac{1}{64} \, , \nonumber \\
{\cal T}_s \! \times \! {\cal T}_s &=& \frac{1152 \pi ^2 \alpha _s^2}{s^2} \left[s^2-(t-u)^2\right] \, ,
\nonumber \\
{\cal T}_t \! \times \! {\cal T}_t &=& -\frac{2304 \pi ^2 \alpha _s^2}{s^2
   \left(m_{\tilde{g}}^2-t\right)^2} \Big\{m_{\tilde{g}}^2 \left[s^2 (t+3 u)+2 s (t^2+2 u^2)+2
   (t+u)^3\right]  \nonumber \\
&& \;\;\;\;\;\;\;\;\;\;\;\;\;\;\;\;\;\;\;\;\;\;   + \, m_{\tilde{g}}^4 \left[s^2-2 s (t+2 u)-6 (t+u)^2\right]+2 m_{\tilde{g}}^6 \left[s+4 (t+u)\right]  \nonumber \\
&& \;\;\;\;\;\;\;\;\;\;\;\;\;\;\;\;\;\;\;\;\;\;  - \, 4 m_{\tilde{g}}^8-t u \left[s^2+2 s u+2 (t^2+u^2)\right]\Big\} \, , \nonumber \\
{\cal T}_s \! \times \! {\cal T}_t &=& -\frac{576 \pi ^2 \alpha _s^2}{s^2 \left(m_{\tilde{g}}^2-t\right)} \left[s (t-u) \left(4 m_{\tilde{g}}^2-t+u\right)+s^3+s^2(u-t)+(t-u)^3\right] \, , \nonumber \\
{\cal T}_t \! \times \! {\cal T}_u &=& -\frac{2304 \pi ^2 \alpha _s^2}{s^2 \left(m_{\tilde{g}}^2-t\right) \left(m_{\tilde{g}}^2-u\right)} \left(m_{\tilde{g}}^4-t u\right) \left[-4 (t+u) m_{\tilde{g}}^2+8
   m_{\tilde{g}}^4+(t-u)^2\right] \, , \nonumber 
\end{eqnarray}
and ${\cal T}_s \! \times \! {\cal T}_u$ and ${\cal T}_u \! \times \! {\cal T}_u$ are related to ${\cal T}_s \! \times \! {\cal T}_t$ and ${\cal T}_t \! \times \! {\cal T}_t$, respectively, by exchanging $t \leftrightarrow u$ in the corresponding expressions.

\subsection*{$\tilde{g} \tilde{g} \to q_A \bar{q}_B$, $\tilde{g} \chi^0_i \to q_A \bar{q}_B$, $\tilde{g} \chi^+_j \to u_A \bar{d}_B$}

These three processes all have $t$- and $u$-channel squark-exchange diagrams, and
$\tilde{g} \tilde{g} \to q_A \bar{q}_B$ also has an $s$-channel gluon-exchange diagram,
whereas the other two processes do not (hence ${\cal T}_s \! \times \! {\cal T}_s = {\cal T}_s \! \times \! {\cal T}_t = {\cal T}_s \! \times \! {\cal T}_u = 0$ for them). Apart from the couplings, these three processes have the same structures 
as for ${\cal T}_t \! \times \! {\cal T}_t$, ${\cal T}_u \! \times \! {\cal T}_u$ and ${\cal T}_t \! \times \! {\cal T}_u$. 
In the case of $\tilde{g} \chi^+_j \to u_A \bar{d}_B$, because the quark CKM 
matrix is involved in the chargino-quark-squark vertex, the indices $A$ and $B$ can be different even if we 
restrict to the case of no generation mixing with only left-right mixing in the third generation for the up-type and down-type squarks. Therefore, it is convenient to write a $6 \times 6$ up-type squark mixing matrix, $Z^{\tilde{U}}$, which relates the interaction eigenstates and mass eigenstates of the up-type squarks as follows:
\begin{eqnarray}
\left(
\begin{array}{c}
 \tilde{u}_L \\
 \tilde{c}_L \\
 \tilde{t}_L \\
 \tilde{u}_R \\
 \tilde{c}_R \\
 \tilde{t}_R \\
\end{array}
\right) = \left(
\begin{array}{cccccc}
 1 & 0 & 0 & 0 & 0 & 0 \\
 0 & 0 & 1 & 0 & 0 & 0 \\
 0 & 0 & 0 & 0 & \cos \theta _{\tilde{t}} & -\sin \theta _{\tilde{t}}
   \\
 0 & 1 & 0 & 0 & 0 & 0 \\
 0 & 0 & 0 & 1 & 0 & 0 \\
 0 & 0 & 0 & 0 & \sin \theta _{\tilde{t}} & \cos \theta _{\tilde{t}} \\
\end{array}
\right) \left(
\begin{array}{c}
 \tilde{u}_1 \\
 \tilde{u}_2 \\
 \tilde{c}_1 \\
 \tilde{c}_2 \\
 \tilde{t}_1 \\
 \tilde{t}_2 \\
\end{array}
\right) \, ,
\end{eqnarray}
where  $\theta _{\tilde{t}}$ is the stop left-right mixing angle. The mass eigenvalues are correspondingly defined as $m_{{\tilde{U}}_1} = m_{\tilde{u}_1}$, $m_{{\tilde{U}}_2} = m_{\tilde{u}_2}$, $m_{{\tilde{U}}_3} = m_{\tilde{c}_1}$, $m_{{\tilde{U}}_4} = m_{\tilde{c}_2}$, $m_{{\tilde{U}}_5} = m_{\tilde{t}_1}$ and $m_{{\tilde{U}}_6} = m_{\tilde{t}_2}$. A similar mixing matrix, $Z^{\tilde{D}}$, is introduced for the down-type squarks, by changing $\theta _{\tilde{t}}$ to the sbottom left-right mixing angle, $\theta _{\tilde{b}}$. The mass eigenvalues $m_{{\tilde{D}}_{1-6}}$ are also defined similarly.  

For $\tilde{g} \tilde{g} \to q_A \bar{q}_B$ we find
\begin{eqnarray}
{\cal T}_s \! \times \! {\cal T}_s &=& \frac{384 \pi ^2 \delta _{\text{AB}} \alpha _s^2}{s^2} \left[2 m_1^2 \left(2 m_3^2+s-t-u\right)+2 m_3^2
   (s-t-u)+2 m_1^4+2 m_3^4+t^2+u^2\right] \, , \nonumber \\
{\cal T}_s \! \times \! {\cal T}_t &=& \sum_{p=1}^6 \frac{192 \pi ^2 \delta _{\text{AB}} \alpha _s^2}{s \left(t-m_{\tilde{f}_p}^2\right)} \Big\{\left[m_1^2 \left(2 m_3^2+s-2
   t\right)+m_3^2 (s-2 t)+m_1^4+m_3^4+t^2\right] \left(|Z_{(A+3) p}^{\tilde{f}}|^2 + |Z_{A p}^{\tilde{f}}|^2 \right) \nonumber \\  
&& \;\;\;\;\;\;\;\;\;\;\;\;\;\;\;\;\;\;\;\;    - \, 2 m_1 m_3 \left(3 m_1^2+3 m_3^2-2
   t-u\right)  \left[Z_{(A+3) p}^{\tilde{f}} \left(Z_{A p}^{\tilde{f}}\right){}^* + Z_{A p}^{\tilde{f}} \left(Z_{(A+3) p}^{\tilde{f}}\right){}^* \right] \Big\}  \, , \nonumber 
\end{eqnarray}
where $m_1 = m_{\tilde{g}}$ and $m_3 = m_{f_A}$. The index $\tilde{f} = \tilde{U}, \tilde{D}$, the index $f = U, D$. $m_{U_{1,2,3}} = m_{u,c,t}$, $m_{D_{1,2,3}} = m_{d,s,b}$, and ${\cal T}_s \! \times \! {\cal T}_u$ is related to ${\cal T}_s \! \times \! {\cal T}_t$ by exchanging $t \leftrightarrow u$. 

For all three processes, ${\cal T}_t \! \times \! {\cal T}_t$, ${\cal T}_u \! \times \! {\cal T}_u$ and ${\cal T}_t \! \times \! {\cal T}_u$ take the following forms:
\begin{eqnarray}
{\cal T}_t \! \times \! {\cal T}_t &=& \sum_{p,q=1}^6 \frac{\pi ^2 c^{\text{tt}} \alpha _s^2}{\left(t-m_{t_p}^2\right)
   \left(t-m_{t_q}^2\right)} \left[2 m_1 m_3 \left(f_L^t(A,p)
   f_R^t(A,q){}^*+f_R^t(A,p) f_L^t(A,q){}^*\right) \right. \nonumber \\ 
&& \;\;\;\;\;\;\;\;\;\;\;\;\;\;\;\;\;\;\;\;\;\;\;\;\;\;\;\;\;\;\;\;\;  \left.  +\left(m_1^2+m_3^2-t\right) \left(f_L^t(A,p)
   f_L^t(A,q){}^*+f_R^t(A,p) f_R^t(A,q){}^*\right)\right] \nonumber \\  
&& \;\;\;\;\;\;\;\;\;\;\;\;\;\;\;\;\;\;\;\;\;\;\;\;\;\;\;\;\;\;\;\;\;\;\; \times \left[2 m_2 m_4 \left(g_L^t(B,p)
   g_R^t(B,q){}^*+g_R^t(B,p) g_L^t(B,q){}^*\right) \right. \nonumber \\ 
&& \;\;\;\;\;\;\;\;\;\;\;\;\;\;\;\;\;\;\;\;\;\;\;\;\;\;\;\;\;\;\;\;\;  \left. +\left(m_2^2+m_4^2-t\right) \left(g_L^t(B,p)
   g_L^t(B,q){}^*+g_R^t(B,p) g_R^t(B,q){}^*\right)\right] \, , \nonumber \\
{\cal T}_t \! \times \! {\cal T}_u &=& - \sum_{p,q=1}^6  \pi ^2 c^{\text{tu}} \alpha _s^2 \Bigg\{ \frac{1}{\left(u-m_{u_p}^2\right) \left(t-m_{t_q}^2\right)}  \nonumber \\ 
&& \!\!\!\!\!\!\!\!\!\!\!\!\!\!\!\!\!\!\!\!\!\!\!\!\!\!\!  \times \bigg\{ \left(f_L^t(A,q){}^* f_L^u(A,p) \Big\{m_2
   g_R^t(B,q){}^* \left[m_4 \left(m_1^2+m_3^2-t\right) g_L^u(B,p)+m_1
   \left(m_1^2+m_2^2-t-u\right) g_R^u(B,p)\right]  \right. \nonumber \\
&& \left. + \, g_L^t(B,q){}^* \left[\left(m_1^2 m_2^2+m_3^2
   m_4^2-t u\right) g_L^u(B,p)+m_1 m_4 \left(m_2^2+m_3^2-u\right) g_R^u(B,p)\right]\Big\} \right. \nonumber \\
&& \left.  + \, m_3 f_L^t(A,q){}^* 
   f_R^u(A,p) \Big\{m_2 g_L^t(B,q){}^* \left[\left(m_1^2+m_4^2-u\right) g_L^u(B,p)+2 m_1 m_4
   g_R^u(B,p)\right] \right. \nonumber \\
&& \left. + \, g_R^t(B,q){}^* \left[m_4 \left(m_3^2+m_4^2-t-u\right) g_L^u(B,p)+m_1
   \left(m_2^2+m_4^2-t\right) g_R^u(B,p)\right]\Big\}\right)  \nonumber \\
&& + \, (L \leftrightarrow R) \bigg\} + (t \leftrightarrow u, m_1 \leftrightarrow m_2) \Bigg\} \, , \nonumber
\end{eqnarray}
where the $(L \leftrightarrow R)$ in ${\cal T}_t \! \times \! {\cal T}_u$ applies to all the $L$ and $R$ in the indices, and the $(t \leftrightarrow u)$ applies to both the $t$ and $u$ in the indices and the Mandelstam variables. Again,
${\cal T}_u \! \times \! {\cal T}_u$ is related to ${\cal T}_t \! \times \! {\cal T}_t$ by exchanging $m_1 \leftrightarrow m_2$ and $t \leftrightarrow u$ in both the indices and the Mandelstam variables. 

The couplings and masses involved in the above expressions are listed below. 

For $\tilde{g} \tilde{g} \to q_A \bar{q}_B$:
\begin{eqnarray}
c_{ini} &=& \frac{1}{64} \, , \nonumber \\
c^{\text{tt}} &=& c^{\text{uu}} \;\; = \;\; \frac{256}{3} \,, \nonumber \\
c^{\text{tu}} &=& -\frac{32}{3} \, , \nonumber \\
m_1 &=& m_2 \;\; = \;\; m_{\tilde{g}} \, , \nonumber \\ 
m_3 &=& m_4 \;\; = \;\; m_{f_A} \, , \nonumber \\
m_{t_p} &=& m_{u_p} \;\; = \;\; m_{\tilde{f}_p} \, , \nonumber \\
f_L^t(A,p) &=& f_L^u(A,p) \;\; = \;\; Z_{(A+3) p}^{\tilde{f}} \, , \nonumber \\
f_R^t(A,p) &=& f_R^u(A,p)\;\; = \;\; -Z_{A p}^{\tilde{f}} \, , \nonumber \\
g_L^t(B,p) &=& g_L^u(B,p)\;\; = \;\; \left(Z_{B p}^{\tilde{f}}\right){}^* \, , \nonumber \\
g_R^t(B,p) &=& g_R^u(B,p) \;\; = \;\; -\left(Z_{(B+3) p}^{\tilde{f}}\right){}^* \, , \nonumber 
\end{eqnarray}
where the index $\tilde{f} = \tilde{U}, \tilde{D}$, and the index $f = U, D$. 

For $\tilde{g} \chi^0_i \to q_A \bar{q}_B$:
\begin{eqnarray}
c_{ini} &=& \frac{1}{8} \, , \nonumber \\
c^{\text{tt}} &=& c^{\text{uu}} \;\; = \;\; - c^{\text{tu}} \;\; = \;\; \frac{8}{\pi  \alpha _s} \, , \nonumber \\
m_1 &=& m_{\tilde{g}} \, , \nonumber \\ 
m_2 &=& m_{\chi^0_i} \, , \nonumber \\ 
m_{t_p} &=& m_{u_p} \;\; = \;\; m_{\tilde{f}_p} \, , \nonumber \\ 
f_L^t(A,p) &=& Z_{(A+3) p}^{\tilde{f}} \, , \nonumber \\
f_R^t(A,p) &=& -Z_{A p}^{\tilde{f}} \, , \nonumber \\ 
g_L^t(B,p) &=& -\frac{i g_2 m_{f_B} c_L^t \left(Z_{(B+3) p}^{\tilde{f}}\right){}^*}{\sqrt{2} m_w}-i \sqrt{2}
   g_2 \left(Z_{B p}^{\tilde{f}}\right){}^* \left[\left(N_{\text{i1}}\right){}^* \tan
   \left(\theta _w\right) \left(Q_{f_B}-T_{f_B}^3\right)+T_{f_B}^3
   \left(N_{\text{i2}}\right){}^*\right] \, , \nonumber \\ 
g_R^t(B,p) &=& i \sqrt{2} g_2 N_{\text{i1}} Q_{f_B} \tan
   \left(\theta _w\right) \left(Z_{(B+3) p}^{\tilde{f}}\right){}^*-\frac{i g_2 m_{f_B} c_R^t
   \left(Z_{B p}^{\tilde{f}}\right){}^*}{\sqrt{2} m_w} \, , \nonumber \\
f_L^u(A,p) &=& i \sqrt{2} g_2 Q_{f_A}
   \left(N_{\text{i1}}\right){}^* \tan \left(\theta _w\right) Z_{(A+3) p}^{\tilde{f}}-\frac{i g_2
   m_{f_A} c_L^u Z_{A p}^{\tilde{f}}}{\sqrt{2} m_w} \, , \nonumber \\ 
f_R^u(A,p) &=& -\frac{i g_2 m_{f_A} c_R^u
   Z_{(A+3) p}^{\tilde{f}}}{\sqrt{2} m_w}-i \sqrt{2} g_2 Z_{A p}^{\tilde{f}} \left[N_{\text{i1}}
   \tan \left(\theta _w\right) \left(Q_{f_A}-T_{f_A}^3\right)+N_{\text{i2}}
   T_{f_A}^3\right] \, , \nonumber \\  
g_L^u(B,p) &=& \left(Z_{B p}^{\tilde{f}}\right){}^* \, , \nonumber \\  
g_R^u(B,p) &=& -\left(Z_{(B+3) p}^{\tilde{f}}\right){}^* \, , \nonumber
\end{eqnarray}
where $N$ is the $4 \! \times \!4$ neutralino mixing matrix as defined in~\cite{Gunion:1984yn}, 
and $g_2$ is the Standard Model $\text{SU(2)}_L$ coupling constant. 
For up-type quark final states, the index $\tilde{f} = \tilde{U}$, and 
\begin{eqnarray}
T_{f_A}^3 &=& T_{f_B}^3 \;\;=\;\; \frac{1}{2} \, , \nonumber \\ 
Q_{f_A} &=& Q_{f_B} \;\;=\;\; \frac{2}{3} \, , \nonumber \\ 
c_L^t &=& c_L^u \;\;=\;\; \left(c_R^t\right){}^* \;\;=\;\; \left(c_R^u\right){}^* \;\;=\;\; \csc (\beta ) \left(N_{\text{i4}}\right){}^* \, , \nonumber \\
m_3 &=& m_{U_A} \, , \nonumber \\
m_4 &=& m_{U_B} \, , \nonumber \\
m_{f_A} &=& m_{U_A} \, , \nonumber \\
m_{f_B} &=& m_{U_B} \, . \nonumber
\end{eqnarray}
$\tb \equiv v_2/v_1$, and the vacuum expectation values of the two Higgs doublets are defined as,
\beq 
\langle H_1 \rangle \equiv \left( \begin{array}{c} v_1 \\ 0 \end{array} \right) ,\;\;
\langle H_2 \rangle \equiv \left( \begin{array}{c} 0 \\ v_2 \end{array} \right) .\;\;
\nonumber
\eeq
For down-type quark final states, the index $\tilde{f} = \tilde{D}$, and 
\begin{eqnarray}
T_{f_A}^3 &=& T_{f_B}^3 \;\;=\;\; - \frac{1}{2} \, , \nonumber \\ 
Q_{f_A} &=& Q_{f_B} \;\;=\;\; - \frac{1}{3} \, , \nonumber \\ 
c_L^t &=& c_L^u \;\;=\;\; \left(c_R^t\right){}^* \;\;=\;\; \left(c_R^u\right){}^* \;\;=\;\; \sec (\beta ) \left(N_{\text{i3}}\right){}^* \, , \nonumber \\
m_3 &=& m_{D_A} \, , \nonumber \\
m_4 &=& m_{D_B} \, , \nonumber \\
m_{f_A} &=& m_{D_A} \, , \nonumber \\
m_{f_B} &=& m_{D_B} \, . \nonumber
\end{eqnarray}

For $\tilde{g} \chi^+_j \to u_A \bar{d}_B$:
\begin{eqnarray}
c_{ini} &=& \frac{1}{8} \, , \nonumber \\
c^{\text{tt}} &=& c^{\text{uu}} \;\; = \;\; - c^{\text{tu}} \;\; = \;\; \frac{8}{\pi  \alpha _s} \, , \nonumber \\
m_1 &=& m_{\tilde{g}} \, , \nonumber \\ 
m_2 &=& m_{\chi^+_j} \, , \nonumber \\ 
m_3 &=& m_{U_A} \, , \nonumber \\ 
m_4 &=& m_{D_B} \, , \nonumber \\ 
m_{t_p} &=& m_{\tilde{U}_p} \, , \nonumber \\
m_{u_p} &=& m_{\tilde{D}_p} \, , \nonumber \\
f_L^t(A,p) &=& Z_{(A+3) p}^{\tilde{U}} \, , \nonumber \\
f_R^t(A,p) &=& -Z_{A p}^{\tilde{U}} \, , \nonumber \\
g_L^t(B,p) &=& \sum_{C=1}^3 \left( \frac{i g_2 \csc (\beta ) K_{\text{CB}} m_{U_C} \left(V_{\text{j2}}\right){}^* \left(Z_{(C+3)
   p}^{\tilde{U}}\right){}^*}{\sqrt{2} m_w}-i g_2 K_{\text{CB}} \left(V_{\text{j1}}\right){}^*
   \left(Z_{C p}^{\tilde{U}}\right){}^* \right) \, , \nonumber \\ 
g_R^t(B,p) &=& \sum_{C=1}^3 \left( \frac{i g_2 \sec (\beta ) K_{\text{CB}}
   U_{\text{j2}} m_{D_B} \left(Z_{C p}^{\tilde{U}}\right){}^*}{\sqrt{2} m_w} \right) \, , \nonumber \\ 
f_L^u(A,p) &=& \sum_{C=1}^3 \left( \frac{i g_2 K_{\text{AC}} \csc (\beta ) m_{U_A} \left(V_{\text{j2}}\right){}^* Z_{C
   p}^{\tilde{D}}}{\sqrt{2} m_w} \right) \, , \nonumber \\ 
f_R^u(A,p) &=& \sum_{C=1}^3 \left( \frac{i g_2 K_{\text{AC}} \sec (\beta )
   U_{\text{j2}} m_{D_C} Z_{(C+3) p}^{\tilde{D}}}{\sqrt{2} m_w}-i g_2 K_{\text{AC}} U_{\text{j1}}
   Z_{C p}^{\tilde{D}} \right) \, , \nonumber \\ 
g_L^u(B,p) &=& \left(Z_{B p}^{\tilde{D}}\right){}^* \, , \nonumber \\
g_R^u(B,p) &=& -\left(Z_{(B+3) p}^{\tilde{D}}\right){}^* \, , \nonumber 
\end{eqnarray}
where the $K$ matrix is the quark CKM matrix, and $U$ and $V$ are the $2 \! \times \!2$ chargino mixing matrices as defined in~\cite{Gunion:1984yn}. 

Finally, we give the s-wave result (i.e., the coefficient $a$ in Eq.~(\ref{sv3})) for the $\tilde{g} \tilde{g} \to q_A \bar{q}_B$ channel, in the limit of a common squark mass and massless quarks, with no generation or left-right mixing in the squark mixing matrices
(the case considered in the main body of the text). In this limit, the contributions from each of the six quark flavor final states are the same, and the result of putting all the six quark flavors together is
\beq
a_{\tilde{g} \tilde{g} \to q_A \bar{q}_B \, \text{limit value}} = \frac{9 \pi  \alpha _s^2 \left(m_{\tilde{g}}^2-m_{\tilde{q}}^2\right){}^2}{8
   m_{\tilde{g}}^2 \left(m_{\tilde{g}}^2+m_{\tilde{q}}^2\right){}^2} \, ,
\label{eq:stucancel}
\eeq
where $m_{\tilde{q}}$ is the common squark mass. When $m_{\tilde{q}} \gg m_{\tilde{g}}$, only the $s$-channel gluon-exchange diagram contributes, and the above expression is proportional to $m_{\tilde{g}}^{-2}$. On the other hand, when $m_{\tilde{q}} \to m_{\tilde{g}}$, the above expression approaches zero. This cancellation of the $s$-, $t$- and $u$-channel contributions results in the feature of the plots at small values of $m_{\tilde{q}}/m_{\tilde{g}}$ that are commented upon in the main body of the text.


\end{document}